\documentclass[a4paper,11pt]{article}
\pdfoutput=1 
             
\usepackage{jheppub_mod} 
 
\newcommand{\al}{\alpha}

\newcommand{\rar}{\rightarrow}
\newcommand{\Rar}{\Rightarrow}

\newcommand{\hal}{\frac{1}{2}}

\newcommand{\del}{\partial}
\newcommand{\mac}{\mathcal}

\newcommand{\iu}{\mathrm{i}}	
\newcommand{\dd}{\mathrm{d}}	
\newcommand{\ee}{\mathrm{e}}	
\newcommand{\N}{\mathbb{N}}	

\newcommand{\te}[1]{\textnormal{#1}}

\graphicspath{{images/}}

\usepackage[sort&compress]{natbib} 

\usepackage{IEEEtrantools} 
\usepackage[font=small]{caption} 
\usepackage{subcaption} 

\usepackage{tikz}
\usetikzlibrary{arrows}
\tikzstyle{spring}=[thick,decorate,decoration={aspect=0.5, segment length=1.5mm, amplitude=1mm,coil}]
\usetikzlibrary{decorations.pathreplacing}
\usetikzlibrary{decorations.pathmorphing}

\title{Modification of the laws of gravity in the DGP model by the presence of a second DGP brane}

\author{Max Warkentin}

\affiliation{Arnold Sommerfeld Center for Theoretical Physics, Ludwig-Maximilians-Universit\"at, \\ Theresienstra{\ss}e 37, 80333 M\"unchen, Germany}

\emailAdd{max.warkentin@physik.uni-muenchen.de}

\abstract{We investigate how the laws of gravity change in the DGP model, if we add a second, parallel 3-brane, endowed with a localized gravitational curvature term. We calculate the gravitational potential energy between two static point sources localized on different branes. We discover a new length scale, which is equal to the geometric mean of the DGP cross-over scale and the separation of the two branes in the extra dimension. For distances, which are larger than this new length scale, we recover the original DGP result, but for smaller distances the gravitational potential is weaker. Furthermore, a region emerges, where a 4-dimensional observer measures a distance independent force. We discuss a possible application of the present scenario for deriving rotation curves of low surface brightness galaxies. Using the Kaluza-Klein description, we observe a curious pattern, in which even and odd KK-modes contribute to the attractive and repulsive parts of the gravitational potential, respectively. Finally, since this setup allows for the existence of a sector of particle species that are interacting arbitrarily weakly with "our" sector, we discuss the implications of this phenomenon for black holes and the bound on the number of species. We find that the behavior is qualitatively different from theories with a normalizable zero-mode graviton.}

\begin{document} 
\maketitle
\flushbottom

\section{Introduction}

We investigate a particular setup in the braneworld scenario, where two flat (tensionless), parallel 3-branes are embedded in a 4+1-dimensional Minkowski spacetime, where the extra dimension is infinite. Apart from the addition of the 2nd brane,\footnote{In Ref. \cite{Padilla2004a} a similar setup with two parallel branes has been investigated. However, there the extra dimension is compact.} the scenario, which we are considering here, is the same as in Ref. \cite{dvali20004d}, where the original DGP model was proposed. In the DGP model, there is a 5-dimensional (5-d) bulk theory, consisting of the 5-d Einstein-Hilbert action with the fundamental Planck mass $M_*$,  and an embedded, tensionless 3-brane, which is endowed with a 4-d Einstein-Hilbert action with the observed Planck mass $M_{\text{P}}$: 
\begin{equation}
S = M_*^3 \int \dd^4 x \, \dd y \, \sqrt{|G|} \mac{R}_5 + M_{\textnormal{P}}^2  \int \dd^4 x \, \sqrt{|g|} R , \label{eq:DGP_action}
\end{equation}
where $y$ is the coordinate of the extra dimension, $|G|$ is the determinant of the bulk metric, $\mac{R}_5$ is the bulk Ricci scalar and $|g|$ (with $g_{\mu \nu}(x^{\mu})=G_{A B} (x^{\mu},y=0)$) and $R$ are the respective quantities on the brane.
Due to the localized curvature term (the second term in \eqref{eq:DGP_action}), the phenomenology of the DGP model is quite distinctive: Gravity behaves as 5-dimensional for distances above the cross-over scale $r_c \equiv \frac{M_{\te{P}}^2}{M_*^3}$, but changes the regime to 4-dimensional behavior for distances below $r_c$. We focus on the gravity part of the model, although a localized matter action in \eqref{eq:DGP_action} is assumed, which contains the Standard Model (SM) fields.

Setups with parallel branes are interesting in several respects: One example is brane inflation \cite{dvali1999brane}. In that scenario, inflation in our universe could have been caused by a relative motion between branes. Another example was given in Ref. \cite{Dvali:1999gf}, where it was suggested that (anti-)baryons could be transported to such a parallel brane leading to a new mechanism of baryogenesis on our brane. 

The goal of the present paper is to show how the physical implications of the DGP model are modified, if there exists a 2nd brane with a localized curvature term. In order to study this, we will use a somewhat simplified model, explained in detail in Section \ref{sec:setup}, and calculate the gravitational potential energy between two static point sources localized on the opposite, parallel branes (in Section \ref{sec:5_dim_potential}). We will find that the 2nd brane enhances the effects of the DGP model by further weakening the 5-d gravity at certain distances.\footnote{In the original DGP setup \cite{dvali20004d,Dvali2001b} it has been found that the localized curvature term shields the short distance physics on the brane from the strong gravity of the bulk.} Furthermore, we will discover that the resulting gravitational potential gives rise to a potentially, phenomenologically interesting new distance independent force in our universe.

Typically, in the context of extra dimensional models, the Kaluza-Klein (KK) language is adopted, where the presence of the (geometrical) extra dimension is traded for the presence of a tower of KK-modes, seen by a 4-d observer. Although such a viewpoint is usually used in the case of a compact extra dimension (since in that case, one gets a discrete mass spectrum), in Section \ref{sec:KK_potential} we will also make use of this KK language and re-derive our results obtained in the fully 5-d treatment. Besides encountering some interesting features, this equivalent viewpoint will serve as a cross-check for our main result.

Finally, we wish to address the question, what are the implications on the present scenario coming from black hole (BH) physics. For general relativity (GR), or theories that behave like GR up to some distance $l_*$, it has been shown \cite{Dvali2007,Dvali2007a,Dvali2008} that BH physics (with a BH size smaller than $l_*$) puts a consistency bound on the short distance cutoff of GR, namely 
\begin{equation}
\Lambda \lesssim \frac{M_{\te{P}}}{\sqrt{N}} , \label{eq:N_bound}
\end{equation}
where $\Lambda$ is the gravity cutoff and $N$ is the number of species in the theory.\footnote{As can be seen from Eq. \eqref{eq:N_bound}, in theories with a large number of species $N$, the maximal cutoff can be significantly lower than the Planck scale.} The physical meaning of expression \eqref{eq:N_bound} is that the parameter $\Lambda^{-1}$ marks the lower bound on the scale of breakdown of semi-classical gravity. For example, the Hawking radiation from a black hole of the size smaller than this scale cannot be treated as thermal, even approximately. This is a clear signal that the theory of general relativity requires a UV-completion at distances shorter than $\Lambda^{-1}$.  

Furthermore, it has been shown in the context of an ADD-type model \cite{Dvali2008c} (see also \cite{Dvali2008e,Aryal1986,Achucarro1995}) 
that several setups with multiple branes, where a BH does not intersect with all of the branes, are classically not static. It was shown that there is a classical time scale, until which the BH will "accrete" all of the other branes.\footnote{In case the time scale is larger than the evaporation time, the BH will evaporate first.} According to Ref. \cite{Dvali2008c}, in the language of species, this can be understood as a "democratization" process for the BH in the following sense: A semi-classical, thermal, BH should evaporate into all species "democratically" (up to greybody factors). However, as long as the BH does not intersect certain branes, it cannot evaporate into the species localized on them. So the process of brane accretion restores that evaporation universality. In this view, the lack of time-independence in the setup is reflected by the lack of universal evaporation.

However, theories of the form considered in the present work, differ from other extra dimensional models (like the ADD model \cite{arkani1998hierarchy} or the RS model \cite{Randall1999,Randall1999a}) in a very important aspect: they modify gravity at large distances, while coinciding with GR at short distance scales. Moreover, as opposed to the above outlined "accretion" scenario, where the tension of the branes is responsible for the attraction between the BH and the branes, in our setup the branes are tensionless. Nevertheless, we will show that there would be an (repulsive) interaction between a BH and the branes. We can, however, switch that interaction off by sending the distance between the branes to infinity. This is not possible in the ADD model, because in that case we would need to send the compactification radius to infinity as well, which would make 4-d gravity to vanish. So how does our scenario fit into the above species picture? What, if any, information do we get from BH physics? We shall address these issues in the last section of this paper and investigate the implications for the bound on the number of species.

\section{Two parallel DGP branes} \label{sec:setup}

As was explained in Ref. \cite{dvali20004d}, in order to see the essential features of the DGP scenario, it is enough to consider a toy model with a bulk scalar field and the respective kinetic term(s) localized on the brane(s). The full theory, involving a spin-2 particle, will add a tensor-structure to the scalar field propagator, but the essential results of the present work will be unaffected.\footnote{\label{fn:scalar_vs_tensor} Since we only consider the interaction between static sources, the full propagator (including the spin-2 and the spin-0 part) would modify our result by an $\mac{O}(1)$ numerical factor.} Hence, for clarity of presentation, we will exclusively deal with the simplified scalar field theory in this paper.

The main contribution of the present work is to discuss a setup in the DGP braneworld scenario, where an additional 3-brane is added, which is parallel (with respect to the extra dimension) to the one representing our universe. The two branes are separated by a distance $R$ along the 5th dimension. In particular, we will consider the following scalar field theory:
\begin{equation}
S = \int \mbox d^4 x \, \dd y \, \left\{ \hal \left( \del_A \phi \right)^2 +  r_c \left[ \delta \left(y \right) + \delta \left(y-R \right) \right] \hal \left( \del_{\mu} \phi \right)^2 + J(x^{\mu},y) \phi \right\}  , \label{eq:5dim_action}
\end{equation}
where $\phi \equiv \phi(x^{\mu},y)$ is a massless scalar field with $A \in \{0,1,2,3,5 \}$ and $\mu \in \{0,1,2,3 \}$. We defined $r_c \equiv \frac{M_{\te{P}}^2}{M_*^3}$, the so-called cross-over length scale, which quantifies the relative strengths of the scalar field propagators on the brane and the bulk, respectively. This simulates the relative strengths of the 4-d and the 5-d gravity in our toy model.

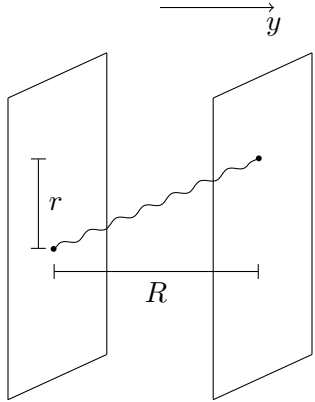
\begin{figure}[t]
\centering
\begin{tikzpicture}
\draw (-2,-2) -- (-2,2);
\draw (-2,2) -- (-0.7,2.6) ;
\draw  (-0.7,2.6) -- (-0.7,-1.4) ;
\draw (-0.7,-1.4) -- (-2,-2);

\draw (0.7,-2) -- (0.7,2);
\draw (0.7,2) -- (2,2.6) ;
\draw  (2,2.6) -- (2,-1.4) ;
\draw (2,-1.4) -- (0.7,-2);

   \draw[->] (0,3.2) -- (1.5,3.2) node[anchor=north] {$y$} ;
   
   \fill[black] (-1.4,0) circle (1.1pt) ;
\fill[black] (1.3,1.2) circle (1.1pt) ;
\draw[decorate,decoration={snake,amplitude=.4mm,segment length=4mm}] (-1.4,0) -- (1.3,1.2);
      
      \draw[|-|] (-1.6,0) -- (-1.6,1.2);
      \draw (-1.6,0.6) node[anchor=west] {$r$};
      
       \draw[|-|] (-1.4,-0.3) -- (1.3,-0.3) ;
        \draw (-0.05,-0.3) node[anchor=north] {$R$};
      
\end{tikzpicture}

\caption{Potential energy between two static point sources on different, parallel branes, mediated by the scalar field.}
\label{fig:branes}
\end{figure}

We want to calculate the potential energy ("would-be" gravitational potential energy) between two static point sources localized on the different branes, as shown in Figure~\ref{fig:branes}.

In the next section we will perform the 5-d calculation, while in the following section we will integrate out the extra dimension and use the KK language.

\section{Potential energy in the 5-d description} \label{sec:5_dim_potential}

The action \eqref{eq:5dim_action} leads to the equation of motion
\begin{equation*}
 \left( \Box - \del_y^2 + r_c \delta (y) \Box + r_c \delta(y-R) \Box \right) \phi  = J(x^{\mu},y) ,
\end{equation*}
with $\Box \equiv \del_{\mu} \del^{\mu}$. 

If the source is static, we can look for static solutions. For this we can first solve
\begin{equation*}
\left( \Delta + \del_y^2+r_c \delta (y) \Delta + r_c \delta(y-R) \Delta  \right) G(\vec{x},\vec{x'};y,y') = - \delta^{(3)}(\vec{x}-\vec{x'} ) \delta (y-y') ,
\end{equation*}
where $\Delta$ is the Laplace operator and $G(\vec{x},\vec{x'};y,y')$ is the Green's function. Although the operator on the left hand side is not translationally invariant along the $y$-direction, we are only interested in the setup, where one of the sources is located on the left brane at $y'=0$ (and $\vec{x'}=0$), so the problem reduces to
\begin{equation*}
\left( \Delta + \del_y^2+r_c \delta (y) \Delta + r_c \delta(y-R) \Delta  \right) G(\vec{x},y) = - \delta^{(3)}(\vec{x}) \delta (y)  .
\end{equation*}
We can Fourier transform this with respect to the coordinates $\vec{x}$ to get
\begin{equation*}
\left( k^2-\del_y^2+r_c \delta (y) k^2 + r_c \delta(y-R) k^2 \right) \widetilde{G}(k,y) = \delta(y),
\end{equation*}
where $k \equiv |\vec{k}|$. Next, we can also Fourier transform with respect to the coordinate of the extra dimension $y$ and find the formal solution
\begin{equation*}
\widetilde{G}_k (k,k^5) = \frac{1- r_c k^2 \left(\widetilde{G}(k,y=0)+ \widetilde{G}(k,y=R) \, \ee^{- \iu k^5 R} \right)}{k^2 + (k^5)^2} ,
\end{equation*}
where $\widetilde{G}_k (k,k^5)$ is the fully Fourier transformed Green's function and $k^5$ is the momentum in the 5th dimension. Transforming back to the $y$-coordinates and using 
\begin{equation*}
\int \frac{\dd k^5}{2 \pi} \frac{\ee^{\iu k^5 y}}{k^2 + (k^5)^2} = \frac{\ee^{-k |y|}}{2 k} ,
\end{equation*}
we get
\begin{equation*}
\widetilde{G}(k,y)=\frac{\ee^{-k |y|}}{2 k}-\frac{r_c k^2 \widetilde{G}(k,0)}{2 k} \ee^{-k |y|} - \frac{r_c k^2 \widetilde{G}(k,R)}{2 k} \ee^{-k |y-R|} . 
\end{equation*}
We can now solve for the coefficients $\widetilde{G}(k,0)$ and $\widetilde{G}(k,R)$ and arrive at
\begin{equation}
\widetilde{G}(k,y)=\frac{\ee^{-k |y|}}{k} \frac{2+r_c k}{(2+r_c k)^2-r_c^2 k^2 \, \ee^{-2 k R}} - r_c \frac{\ee^{-k |y-R|} \, \ee^{-k R}}{(2+r_c k)^2-r_c^2 k^2 \, \ee^{-2 k R}} . \label{eq:propagator}
\end{equation} 
For 
\begin{equation}
J(x^{\mu},y)=g \left[ \delta^{(3)}(\vec{x}) \delta (y)+\delta^{(3)}(\vec{x}-\vec{r}) \delta (y-R) \right] \label{eq:source}
\end{equation} 
(see Figure~\ref{fig:branes}) we find as the potential energy between the sources
\begin{equation}
V(r,R) = - g^2 G(\vec{r},R) = - \frac{g^2}{\pi^2} \frac{1}{r} \int_0^{\infty} \dd k \frac{\sin{(kr)} \, \ee^{-k R}}{(2+r_c k)^2-r_c^2 k^2 \, \ee^{-2 k R}} ,  \label{eq:5dim_potential}
\end{equation} 
where $r \equiv |\vec{r}|$. 

We see already from (\ref{eq:5dim_action}), with (\ref{eq:source}), that if the field $\phi$ is to mimic the graviton, the correct coupling constant should be $g \propto M_*^{-3/2}=\frac{\sqrt{r_c}}{M_{\te{P}}}$ (for unit-mass point-sources). We fix the numerical coefficient such that\footnote{With this normalization, in the limit $R \rar 0$, where our system reduces to the original DGP setup, our potential energy will be only half the value. This is easy to see, since for $R \rar 0$, \eqref{eq:5dim_action} reduces to the DGP system with $r_c \rar 2 r_c$. However, with this choice we recover the usual Newton's potential for close enough point sources on our brane, as we will see in Subsection~\ref{sec:force_along_brane}.} 
\begin{equation}
g^2=\frac{r_c}{2 M_{\te{P}}^2} .  \label{eq:fix_coupling}
\end{equation}

The resulting integral cannot be solved exactly, but we can extract the leading order behavior for the three interesting regimes to be specified shortly. Let us rewrite (\ref{eq:5dim_potential}) as
\begin{equation}
V(r,R) = - \frac{g^2}{4 \pi^2} \frac{1}{r r_c} J ,
\end{equation} 
where
\begin{equation}
J = \int_0^{\infty} \dd x \, \frac{\sin{x} \, \ee^{-\frac{R}{r} x}}{\frac{r}{r_c} + x + \frac{1}{4} \frac{r_c}{r} x^2 \left(1- \ee^{-2 \frac{R}{r} x} \right)} . \label{eq:J}
\end{equation} 
We want to approximate the result for the regime $r \ll r_c$ and $R \ll r_c$. Since the exponential cuts the integral off at $\frac{r}{R}$, we can approximate the above integral as
\begin{equation}
J \simeq \int_0^{\frac{r}{R}} \dd x \, \frac{\sin{x}}{\frac{r}{r_c} + x + \hal \frac{\rho^2}{r^2} x^3} ,  \label{eq:first_approx_J}
\end{equation} 
where we have introduced the new length scale 
\begin{equation}
\rho \equiv \sqrt{r_c R}, \label{eq:def_rho}
\end{equation} 
because it will emerge in the result. In the regime $r \gg R$, we can further rewrite
\begin{equation}
J \simeq \int_0^{\infty} \dd x \, \frac{\sin{x}}{\frac{r}{r_c} + x + \hal \frac{\rho^2}{r^2} x^3} .   \label{eq:second_approx_J}
\end{equation}
For $r \gg \rho$, the last term in the denominator of \eqref{eq:second_approx_J} is never dominant in the relevant integration region ($x \lesssim 1$), hence we can approximate
\begin{equation}
J \simeq  \int_0^{\infty} \dd x \, \frac{\sin{x}}{\frac{r}{r_c} + x} = J_{\te{DGP}} + \mac{O} (1) \cdot \ee^{-\sqrt{2} \frac{r}{\rho}}  , \qquad r \gg \rho , \label{eq:approx_J_case_1}
\end{equation}
with
\begin{equation}
J_{\te{DGP}} \equiv  \sin{\left( \frac{r}{r_c} \right)} \mbox{Ci} \left( \frac{r}{r_c} \right) + \cos \left( \frac{r}{r_c} \right) \left[ \frac{\pi}{2} - \mbox{Si} \left(\frac{r}{r_c}\right)  \right] , \label{eq:J_DGP}
\end{equation}
where $\mbox{Si}(z)$ and $\mbox{Ci}(z)$ are the sine integral and the cosine integral, respectively. So in this regime, the potential energy is approximately the same as the one found in the original DGP setup \cite{dvali20004d}.\footnote{ \label{fn:half_potential_energy} This is not surprising, since for two point sources separated at a large distance $r$, their bulk separation becomes less important and the system with two branes behaves like a system with one brane provided that $r_c \rar 2 r_c$.} Since we are only interested in the regime $r \ll r_c$, the above expression reduces to
\begin{equation*}
J_{\te{DGP}} \simeq \frac{\pi}{2}+ \left( \mac{O} (1) + \ln{\frac{r}{r_c}} \right) \frac{r}{r_c} + \mac{O} \left( \frac{r^2}{r_c^2} \right) , \qquad r \ll r_c  . 
\end{equation*}

For $r \ll \rho$ the last term in the denominator of \eqref{eq:second_approx_J} cannot be neglected. However, now we can approximate
\begin{IEEEeqnarray}{rCl} 
J &\simeq & \int_0^{\infty} \dd x \, \frac{\frac{\sin{x}}{x}}{1 + \hal \frac{\rho^2}{r^2} x^2} + \mac{O} \left( \frac{r}{r_c} \right) + \mac{O} \left( \frac{R}{r_c} \right) ,   \nonumber \\ 
&=& \frac{\pi}{2} \left(1- \ee^{- \sqrt{2} \frac{r}{\rho}} \right) + \mac{O} \left( \frac{r}{r_c} \right) + \mac{O} \left( \frac{R}{r_c} \right) , \nonumber \\
&\simeq &  \frac{\pi}{\sqrt{2}}  \frac{r}{\rho} + \mac{O}\left( \frac{r^2}{\rho^2} \right) + \mac{O} \left( \frac{r}{r_c} \right) + \mac{O} \left( \frac{R}{r_c} \right) , \quad  r \ll \rho . 
\label{eq:approx_J_case_2}
\end{IEEEeqnarray} 

Since an asymptotic expansion of the integral \eqref{eq:J} is not available, which would justify the approximations \eqref{eq:approx_J_case_1} and \eqref{eq:approx_J_case_2} analytically, we use numerical means to show in appendix \ref{sec:numerics_different} that the approximations are nevertheless reliable in the stated regimes. We also verify numerically (in appendix \ref{sec:numerics_different}) that the corrections to the leading order terms, given above, are correct. 

Finally, in the regime $r \ll R$, we can approximate $\sin{x} \simeq x$ in the numerator of \eqref{eq:first_approx_J} and (using \texttt{Mathematica} \cite{Mathematica}) find\footnote{The sub-leading terms in this expression were extracted numerically. However, we can understand some of their features from the following consideration. Note that in going from \eqref{eq:J} to \eqref{eq:J_case_3}, we omitted terms with positive (possible fractional) powers of $r/R$ (and additional powers of $R/r_c$), which are subleading in this regime. What is the next-to-leading power of $r/R$? On physical grounds, we know that the force \emph{along the brane} should go to zero for $r \rar 0$. Also, we can see from \eqref{eq:J} explicitly that $\frac{\del}{\del r} (J/r) \rar 0$ for $r \rar 0$. Hence we know that the leading correction (in $r/R$) to \eqref{eq:J_case_3} has to vanish faster than quadratic. From this argument it follows that the (magnitude of the) force in the regime $r \ll R$ is bounded by the (magnitude of the) force in regime $R \ll r \ll \rho$ and will approach zero eventually. However, to determine how fast it will approach zero, we resort to a numerical analysis (see appendix \ref{sec:numerics_different}), which shows that the leading correction in \eqref{eq:J_case_3} is cubic in $r/R$ (with a coefficient of order $R/r_c$).}
\begin{IEEEeqnarray}{rCl} 
J &\simeq & \int_0^{\frac{r}{R}} \dd x \, \frac{x}{\frac{r}{r_c} + x + \hal \frac{\rho^2}{r^2} x^3} , \nonumber \\
&\simeq& \frac{\pi}{\sqrt{2} }\frac{r}{\rho} h \left( \frac{R}{r_c} \right)  + \mac{O} \left( \frac{r^3}{R^3} \right) \cdot \mac{O} \left( \frac{R}{r_c} \right), \qquad r \ll R , \label{eq:J_case_3}
\end{IEEEeqnarray} 
where 
\begin{equation}
h \left( \frac{R}{r_c} \right) \equiv  1 - \frac{1}{\pi \sqrt{2}} \left(\mac{O} (1) + \ln{\frac{R}{r_c}} \right) \sqrt{\frac{R}{r_c}}  + \left(\text{subleading orders of }  \frac{R}{r_c}  \right) . \label{eq:def_corr}
\end{equation}

We can now state the asymptotic behavior of the potential energy in the three stated regimes. For $r, R \ll r_c$
\begin{IEEEeqnarray}{rCl}
&& V(r,R) = - \frac{1}{ 8 \pi^2 M_{\te{P}}^2} \frac{1}{r} J \nonumber \\
&& \simeq - \frac{1}{16 \pi M_{\te{P}}^2} \times \left\{ 
\begin{aligned} 
&\mbox{(I)} & &\frac{1}{r} + \frac{2}{\pi} \left( \mac{O} (1) + \ln{\frac{r}{r_c}} \right) \frac{1}{r_c} + \mac{O} (1) \frac{1}{r} \cdot \ee^{-\sqrt{2} \frac{r}{\rho}} ,  & \rho \ll r& , \\ 
&\mbox{(II)}  & &\left( \sqrt{2} \frac{1}{\rho} - \frac{r}{\rho^2} \right) + \frac{1}{r} \left( \mac{O}\left( \frac{r^3}{\rho^3} \right) + \mac{O} \left( \frac{r}{r_c} \right) + \mac{O} \left( \frac{R}{r_c} \right) \right) , & R \ll r \ll \rho&,  \\
&\mbox{(III)}  & & \sqrt{2}\frac{1}{\rho} h \left( \frac{R}{r_c} \right)  + \frac{1}{r}\mac{O} \left( \frac{r^3}{R^3} \right) \cdot \mac{O} \left( \frac{R}{r_c} \right)    ,   & r \ll R& ,
\end{aligned} \right. \nonumber \\ \label{eq:5dim_potential_approx}
\end{IEEEeqnarray} 
with $\rho$ and $h \left( \frac{R}{r_c} \right)$ as defined in \eqref{eq:def_rho} and \eqref{eq:def_corr}, respectively.

So we see that for cases (II) and (III) the potential is proportional to $\frac{1}{\sqrt{R}}$ (for $R \ll r_c$) and it is weaker than the $\frac{1}{r^2+R^2}$-potential, which we would get without the DGP branes, since $\frac{1}{r_c \rho} \ll \frac{1}{r^2+R^2}$. It is also weaker than the $\frac{1}{D}$-potential, which has been found in Ref. \cite{Dvali2001b},\footnote{There, the setup contains only one DGP brane and the potential energy is calculated between one point source on the brane and a second point source outside the brane. $D$ is the larger one of the distances either along the 3-brane or along the extra dimension.} because $\frac{1}{r_c \rho} \ll \frac{1}{r_c D}$. There, it was argued that the DGP brane acts as a kind of anti-gravity and reduces 5-d gravity to 4-d gravity (see next subsection). Now we find in the present work that two branes enhance this effect and further weaken the 5-d gravity.

\subsection{Screening of the 5-d force by the branes}

In Ref. \cite{Dvali2001b} the propagator for the DGP model (in the presence of one brane) between two points at $y$ and $y_0$ along the extra dimension was calculated. It was approximated in the regime $r \ll r_c$ (and hence $k r_c \gg 1$) as
\begin{equation*}
{}^{\te{(1B)}}\widetilde{G}(k;y,y_0) \cong \frac{1}{k} \ee^{- k |y-y_0|} -\frac{1}{k} \ee^{- k ( |y|+|y_0|)}+\frac{1}{k^2 r_c} \ee^{- k ( |y|+|y_0|)} .
\end{equation*} 
Then, the potential energy between two point sources $m_1$ and $m_2$ at $y$ and $y_0$ is
\begin{IEEEeqnarray}{rCl} 
{}^{\te{(1B)}}V(r;y,y_0) & \sim &  - \frac{m_1 m_2}{M_*^3} \left( \frac{1}{r^2+|y-y_0|^2}-\frac{1}{r^2+( |y|+|y_0|)^2} \right. \nonumber \\
&& +\> \left. \frac{1}{r r_c} \arctan \frac{r}{|y|+|y_0|} \right) . \label{eq:DGP_full_potential}
\end{IEEEeqnarray}

Although this result gives us the potential energy between two point sources, we can interpret it as the potential due to one point source in the presence of a brane, because of the following reason: For a static point source with mass $m_1$ at $(\vec{r},y)$, the solution for the scalar field (evaluated at the point $(\vec{x},y_0)$) is $\phi(\vec{x},y_0) \sim \frac{m_1}{M_*^{3/2}} \int \dd^3 \vec{k} \, \ee^{\iu \vec{k} (\vec{x}-\vec{r})} {}^{\te{(1B)}}\widetilde{G}(k;y,y_0)$. But since our scalar field $\phi$ is essentially the graviton (see footnote~\ref{fn:scalar_vs_tensor}), the gravitational potential (\textit{not} the potential \textit{energy}) is $M_*^{-3/2} \phi$, which is the same as $\frac{1}{m_2} {}^{\te{(1B)}}V(r;y,y_0)$ (from \eqref{eq:DGP_full_potential}). Hence, expression \eqref{eq:DGP_full_potential} can be viewed as the potential energy of a (probe) point mass $m_2$ at $(\vec{x}=\vec{0},y_0)$ due to a point mass $m_1$ at location $(\vec{r},y)$.

We see that the resulting potential differs from a mere $\left( - \frac{m_2}{M_*^3} \frac{1}{r^2 + (y-y_0)^2} \right)$-potential quite dramatically, due to the presence of the DGP-brane. The situation is similar to the so-called image problem from classical electrostatics, so we can use that intuition to interpret the present situation. There, we consider the situation of a point charge in the presence of a perfectly conducting plate. The (positive) point charge induces a negative charge on the plate and the resulting potential is the same as if a mirror image charge has been introduced on the opposite side of the plate. 

Similarly, the potential in the DGP scenario has a form, as if the brane introduced a negative (anti-gravitating) mass on the opposite side of the brane (opposite to the probe mass). For probe masses at $y_0 \le 0$ (with the source $m_1$ at positive $y$) the "image" mass cancels the 5-d potential, while for probe masses at $y_0 > 0$ the image mass enhances the attraction between $m_1$ and $m_2$. 

However, there is an important difference to the situation in the electrostatics case. In the DGP scenario there is still an attractive (but 4-d) potential left (since for $|y|+|y_0| \gg r$ the last term in \eqref{eq:DGP_full_potential} goes to $\frac{1}{r_c} \frac{1}{|y|+|y_0|}$). So, if we consider a point mass in the bulk near the brane, the brane will have a repulsive force on the point mass. 

Now let us return to our two-branes setup. In the last section we found that in the presence of two branes, the attractive potential between two point masses is further reduced. So we can infer that the effect of the branes is to introduce even stronger repulsive (anti-gravitating) point masses. Hence, if we would consider a point mass on our brane, then the parallel brane (if it would be empty of sources) would be repelled.\footnote{Of course, strictly speaking in our present construction the branes are boundaries and hence do not respond to dynamics. However, we can switch gears and consider a situation, where the mechanism, which localizes a point mass on the brane dominates over the mechanism, which fixes the branes to particular spacetime points.} This repulsion in our two-branes setup would be even greater than the repulsion between a brane and a point source in empty space in the original DGP setup.

\subsection{Force along the brane} \label{sec:force_along_brane}

Although the potential \eqref{eq:5dim_potential_approx} is $R$-dependent, the sources in our scenario are, per construction, localized on the brane, so nothing can move into the bulk (except the graviton itself). Put differently, every force, which is orthogonal to the brane, is compensated by the force, which is responsible for localizing the matter on the brane. Hence, in the following subsection we will be interested in the force along the $r$-direction. 

\begin{figure}[t]
\centering
    \includegraphics[height=5cm]{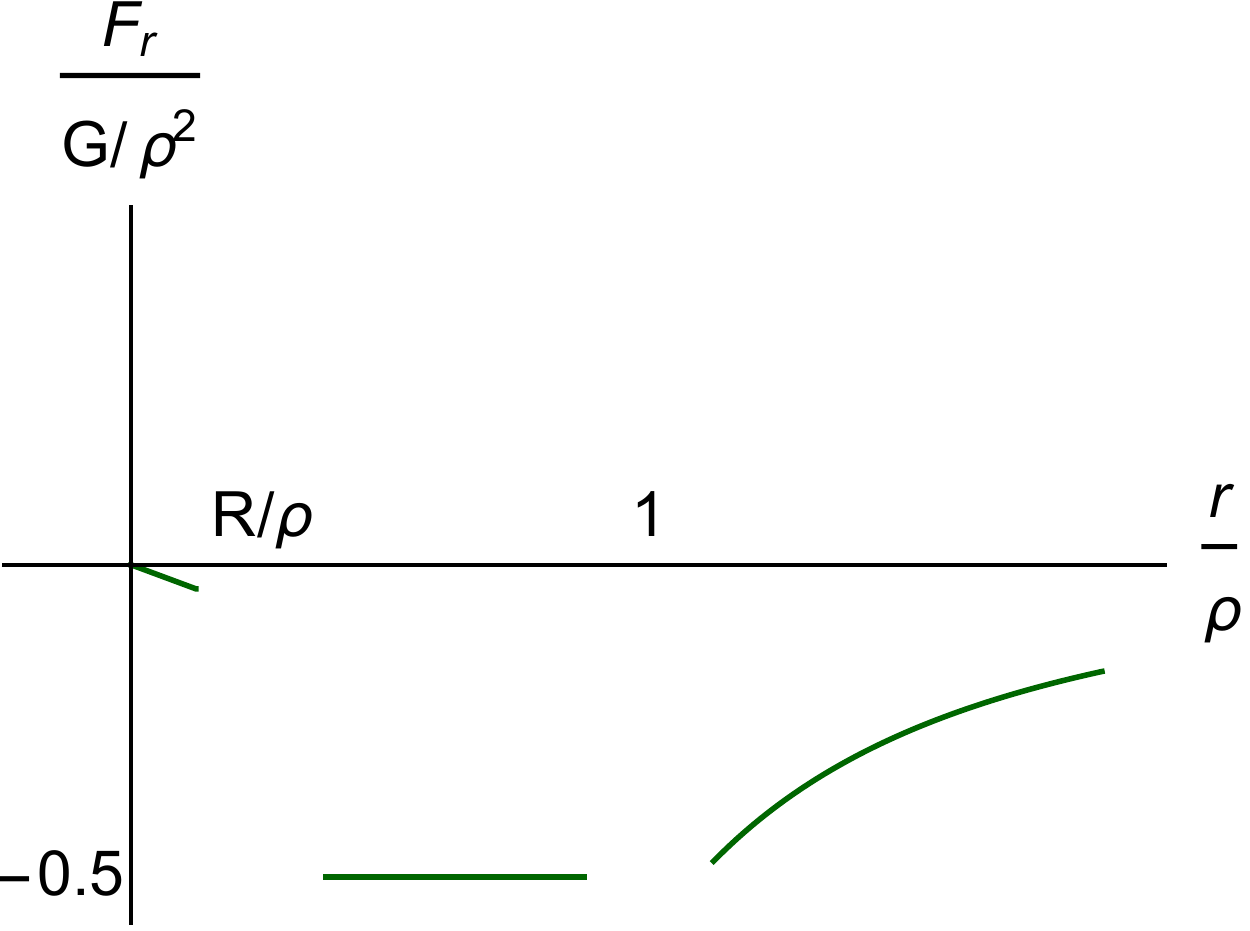}
    \caption[]{This (linear) plot shows the force $F_r$, expressed in units of  $\frac{G}{\rho^2}$, plotted against the distance along the 3-brane, $r$, expressed in units of $\rho$. The force takes different asymptotic forms in the regimes $r \ll R$, $R \ll r \ll \rho$ and $\rho \ll r$.}
    \label{fig:force}
   \end{figure}

We find
\begin{equation}
F_r \equiv - \frac{\del V(r,R)}{\del r} \simeq - \frac{1}{16 \pi M_{\te{P}}^2} \times \left\{ \begin{aligned} 
&\mbox{(I)} & &\frac{1}{r^2} + \mac{O} \left( \frac{1}{r \rho} \right) \cdot \ee^{-\sqrt{2} \frac{r}{\rho}} + \mac{O} \left( \frac{1}{r r_c} \right) , & \rho \ll r& , \\ 
&\mbox{(II)} & &\frac{1}{\rho^2} + \frac{1}{\rho^2} \mac{O} \left( \frac{r}{\rho} \right) + \frac{1}{r^2} \mac{O} \left( \frac{R}{r_c}\right)  , & \quad R \ll r \ll \rho&,  \\
&\mbox{(III)} & & \frac{1}{R^2} \mac{O} \left( \frac{r}{R} \right) \cdot \mac{O} \left( \frac{R}{r_c} \right)  ,  & r \ll R& .
\end{aligned} \right. \label{eq:force}
\end{equation} 
A brane-observer, carrying a static point source, would measure (in the respective asymptotic regimes) a force as sketched in Figure~\ref{fig:force}.\footnote{$G=(8 \pi M_{\te{P}}^2)^{-1}$ is the Newton's constant.}
  
We see that the following picture emerges: Let us assume that we are an observer, who can be approximated by a static point source with mass $m$, living on "our" brane at $y=0$. Now consider there is a different static point source with mass $M$ located at the parallel brane at the location $y=R$ and $\vec{r}=0$.
We start on our brane at $\vec{r}=0$ and probe the gravitational force along the spatial dimensions of our world. If we start increasing the world-volume-distance $r$, we would measure a linearly increasing force, which would be suppressed by $\mac{O} \left( \frac{r}{R} \right)$ (with respect to regime (II)) as long as we stay within the region $r \ll R$. However, if we would measure again at $R \ll r \ll \rho$, we would determine the constant force $F_r = - \frac{G M m}{ 2 \rho^2}$. Finally, for $r \gg \rho$, we would measure the same (or rather 1/2) 4-d force $F_r =- \frac{G M m}{ 2 r^2}$, as if the 2nd point source was located on our brane and there were no extra dimensions. Note that in all of those regimes the force behaves differently than a usual 5-d force would, which would be due to a source on the parallel brane in the absence of the localized kinetic term (DGP-term).

Now, it is interesting that in this scenario a (spatially) constant attractive force emerges beyond some length scale $R$. Let us entertain the possibility for a moment that this force could compensate the decreasing force (as we increase $r$) coming from the baryonic matter on our brane (in our present scenario this baryonic matter would constitute a galaxy, centered around $\vec{r}=0$). Then, one might hope that such a scenario could explain the fact that the gravitational force being exerted from, say, the interior of our galaxy, is stronger than the baryonic mass distribution would imply (see e.g. Ref. \cite{Rubin1980}) without the need of postulating the presence of dark matter on our brane.

In the regime $R \ll r \ll \rho$ the mass $M$ of the source on the 2nd brane has to be much larger than the enclosed baryonic matter on our brane in order to compete with the baryonic Newton's force. So one is forced to take 
\begin{equation}
M \sim M_B(r_*) \left( \frac{\rho}{r_*} \right)^2 \gg  M_B(r_*) , \label{eq:M_M_B}
\end{equation}
where $M_B(r_*)$ denotes the enclosed baryonic mass at the radius $r_*$, at which the baryonic Newton's force starts to decrease. One also has to assume $r_* \sim R$, so that the constant force does not spoil the observations, which are compatible with the usual baryonic Newton's force for $r \lesssim r_*$. 

In this scenario, the constant force would dominate at the outer galactic region and the rotation-velocities of orbiting objects would go like $v(r) \propto \frac{\sqrt{G M}}{\rho} \sqrt{r}$. Although this behavior is not observed in more massive galaxies, where the rotation-curves tend to become flat, in low surface brightness galaxies the circular velocities at large radii are observed to go as $\propto \sqrt{r}$ \cite{Mannheim1997} or even $\propto r$ \cite{Borriello2001}.

\begin{figure}[t]
\centering
    \includegraphics[width=0.7\linewidth]{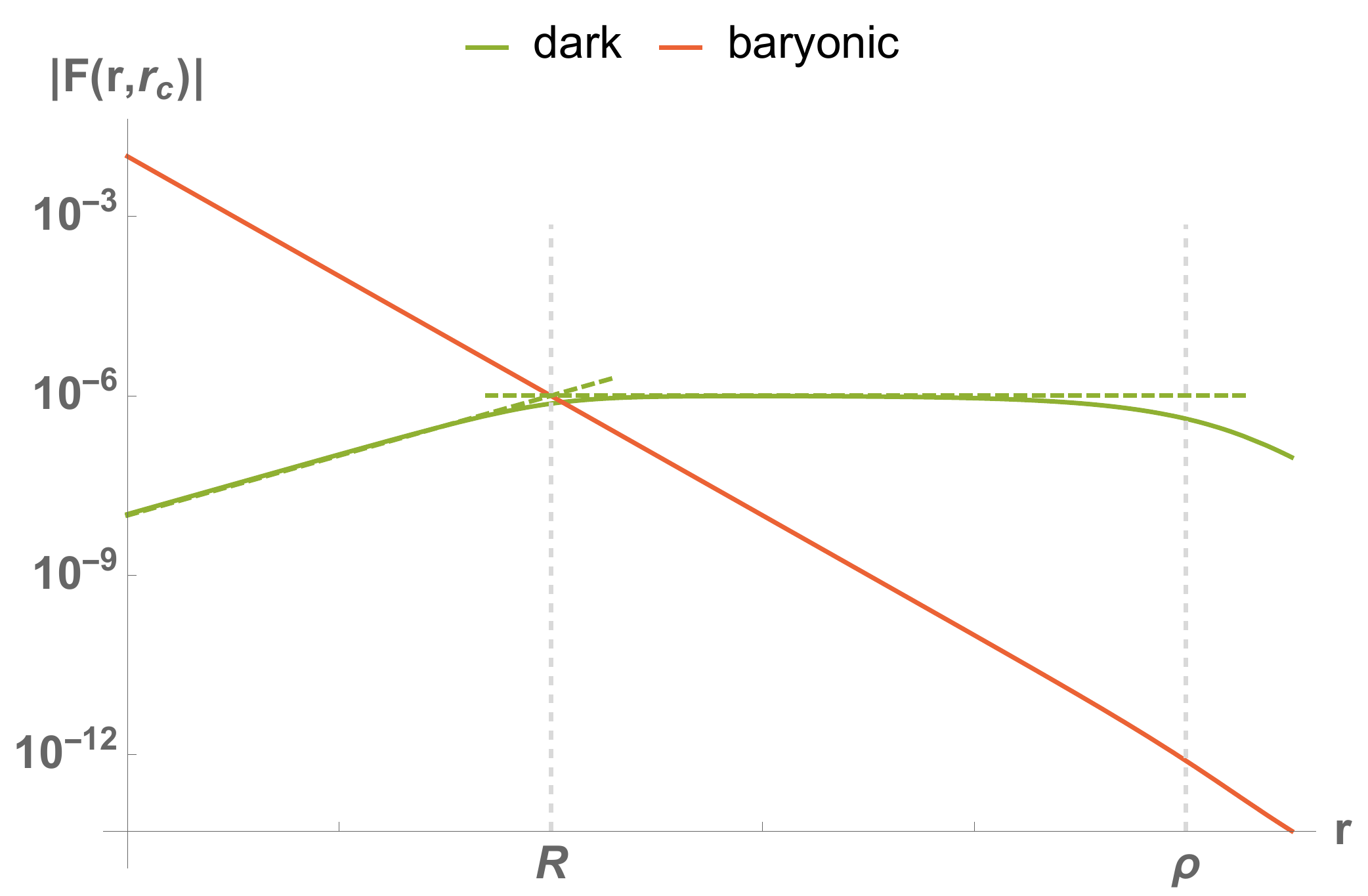}
    \caption[]{ This log-log plot shows the force along the brane as a function of $r$ (in units of $R$). The "dark" force (due to a point source with mass $M$ on the parallel brane), expressed in units of $\frac{G M m}{2 R^2}$ (which is just $f$ from \eqref{eq:f_numerical}), is plotted as the green solid curve. The green dashed curves are its asymptotes in regimes (III) and (II), given by $\frac{r}{r_c}$ and $\frac{1}{r_c}$. The red curve is the "baryonic" force (due to a point source on the same brane), expressed in the same units. The "baryonic" force is obtained numerically from \eqref{eq:potential_baryonic}, with $F_{\te{B}}=-\frac{\partial}{\partial r} V_{\te{B}}$ (see also footnote~\ref{fn:numerical_force}). We did not plot the asymptote $\sim 1/r^2$, because in the regime $r \ll \rho$ it is indistinguishable from $F_{\te{B}}$. The point mass, which gives rise to $F_{\te{B}}$, is chosen such that both forces are of the same order at $r=R$. We chose the ratio $r_c/R=10^6$.}
    \label{fig:galaxy_plot}
   \end{figure}

Note that for the validity of the above outlined scenario we have to check, if the Newton's force between the baryonic matter on our brane and the observer would not be modified by the parallel brane in a severe way. For this, we can again approximate the baryonic matter as a point mass and calculate the potential energy, as given in \eqref{eq:potential_again}, where $r$ is the distance between the point sources. The integral one has to evaluate for this, is even more involved than the one we considered previously. Hence the analysis is carried out numerically and displayed in appendix \ref{sec:numerics_same}. We see (from Figure~\ref{fig:f_baryonic}) that the Newton's force is unchanged for $r \ll \rho$ (regimes (II) and (III)). For $\rho \ll r \ll r_c$, the force crosses over to the same Newton's force, but with one half its value. This is due to the fact that in this regime the observer is at such a large distance that the two branes appear to lie on top of each other and hence $r_c \rar 2 r_c$. Since after $r_*$ the Newton's force falls off, the proposed scenario remains valid. Figure~\ref{fig:galaxy_plot} 
summarizes the above statement by comparing the "baryonic" force to the "dark" force (which is due to the point source on the parallel brane). We see that in this scenario, for $r \ll R$ (the interior of our galaxy), the usual baryonic force is much larger than the dark force. However, for $r \gtrsim R$, the constant dark force takes over.

It should be emphasized that the above distance independent force was derived for a point source at the 2nd brane. So it could be interesting to take an extended source and investigate how the asymptotic rotation-velocities would be impacted. It has been pointed out \cite{Oman2015} that there is a substantial diversity in the rotation curves in low surface brightness galaxies. This diversity might be explained by varying mass distributions on the parallel brane. Since in the present scenario only the parallel brane would contain this "dark" matter distribution, it does not have to be weakly interacting, but could even form a highly localized density distribution. 

Let us estimate, very crudely, the values of the involved length scales, such that the above outlined scenario can be valid. If we wish the distance independent force to be operational at $r \gtrsim 1\,$kpc, we should choose $R \sim 1\,$kpc. If we then want this regime to hold up to at least $r \gtrsim 10^2$ kpc, we should take 
\begin{equation}
\rho \gtrsim 10^2\,\te{kpc}, \label{eq:rho_bound}
\end{equation}
because for $r \gtrsim \rho$ the potential drops and the rotation-velocity would start to decrease. Since there is the constraint $r_c \gtrsim H_0^{-1} \sim 10^6\,$kpc from cosmological observations \cite{Deffayet2001},\footnote{In some surveys \cite{Lombriser2009,Fang2008} 
it is even claimed that cosmological observations exclude a tensionless DGP brane.} we actually need $\rho \gtrsim 10^3\,$kpc, which is significantly larger than the size of the galaxy. This constraint, however, does not necessarily imply that the constant force should extend to galaxy cluster scales, because our analysis shows that the constant force gets corrections as $r$ approaches $\rho$. For example, from Figure~\ref{fig:galaxy_plot}, we see that the force starts to deviate from its constant behavior for $r \gtrsim 10^2\,$kpc, if we take $\rho = 10^3\,$kpc. A more detailed investigation is needed in order to establish the exact behavior of this new force at those large scales and to compare the results to predictions made by the usual cold dark matter scenario and to observations. We will postpone such an investigation to a future work.

Let us finally note that if the above scenario is valid, we have to ask the question, why is the mass on the parallel brane so tightly related to the mass in the "baryonic" galaxy, as seen from \eqref{eq:M_M_B}. This question should then be addressed in the investigation of galaxy formation.

\section{Kaluza-Klein decomposition} \label{sec:KK_potential}

The Green's function and the resulting potential energy can be derived in the Kaluza-Klein language. The advantage of this calculation is that we can gain more insight into the system and compare the results in both languages. 

Let us shift the two branes, so that the system is symmetric around $y=0$:
\begin{equation}
S = \int \mbox d^4 x \, \dd y \, \left\{ \hal \left( \del_A \phi \right)^2 +  r_c \left[ \delta \left(y + R/2 \right) + \delta \left(y-R/2 \right) \right] \hal \left( \del_{\mu} \phi \right)^2 + J(x^{\mu},y) \phi \right\} . \label{eq:action}
\end{equation}
We can now expand the field $\phi$ in the following way:
\begin{equation*}
\phi(x^{\mu},y)= \sum \limits_{\alpha=1}^2 \int_0^{\infty} \dd m \, \psi_{m,\alpha}(y) \phi_{m,\alpha}(x^{\mu}),
\end{equation*}
where $\phi_{m,\alpha}(x^{\mu})$ are the KK-modes and $\psi_{m,\alpha}(y)$ are the mode functions (or wave functions) constituting a complete basis of the $y$-space. We can make the Lagrangian in (\ref{eq:action}) diagonal in $\phi_{m,\alpha}(x^{\mu})$, if the mode functions satisfy the equation
\begin{equation}
\left\{ \del_y^2 + m^2 \left[ 1+r_c \delta \left(y + R/2\right) + r_c \delta \left(y-R/2 \right) \right] \right\} \psi_{m,\alpha}(y) =0 , \label{eq:schroedinger}
\end{equation}
which implies the orthonormality condition
\begin{equation}
\int \dd y \, \psi_{m,\alpha}(y) \psi_{m',\alpha'}(y) \left[ 1+r_c \delta \left(y + R/2\right) + r_c \delta \left(y-R/2 \right) \right] = \delta (m-m') \delta_{\alpha \alpha'} . \label{eq:completeness}
\end{equation}
Then, the action becomes
\begin{IEEEeqnarray}{c} 
S=\int \dd^4 x \, \sum_{\alpha} \int \dd m \, \left\{ - \hal \phi_{m,\alpha}(x^{\mu}) \left[ \Box + m^2 \right] \phi_{m,\alpha}(x^{\mu}) + \phi_{m,\alpha}(x^{\mu}) J_{m,\alpha}(x^{\mu}) \right\} , \label{eq:KK_action}  \nonumber\\*
\end{IEEEeqnarray}
with $J_{m,\alpha}(x^{\mu})= \int \dd y \, J(x^{\mu},y) \psi_{m,\alpha}(y)$.\footnote{Note that $J_{m,\alpha}(x^{\mu})$ is \emph{not} a KK-mode of $J(x^{\mu},y)$, which would rather be $\tilde{J}_{m,\alpha}(x^{\mu}) \equiv \int \dd y \, \psi_{m,\alpha}(y) \left[ 1+r_c \delta \left(y + R/2\right) + r_c \delta \left(y-R/2 \right) \right] J(x^{\mu},y)$.} We can determine the basis $\{\psi_{m,\alpha}(y)\}$ by solving eq. (\ref{eq:schroedinger}). This is done in appendix \ref{sec:mode_functions}.

\begin{figure}[t]
    \centering
    \begin{subfigure}{0.48\textwidth}
        \includegraphics[width=\textwidth]{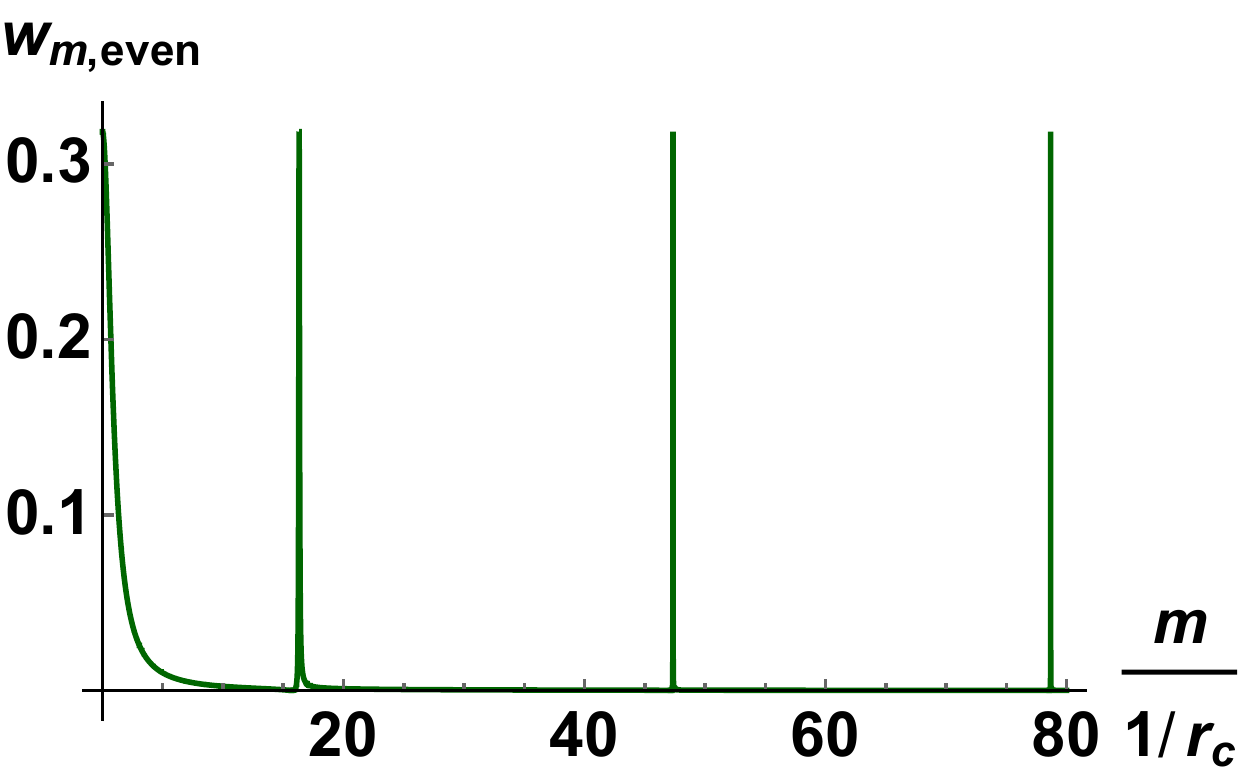}
        \caption{Even}
     \end{subfigure}
    ~ 
    \begin{subfigure}{0.48\textwidth}
        \includegraphics[width=\textwidth]{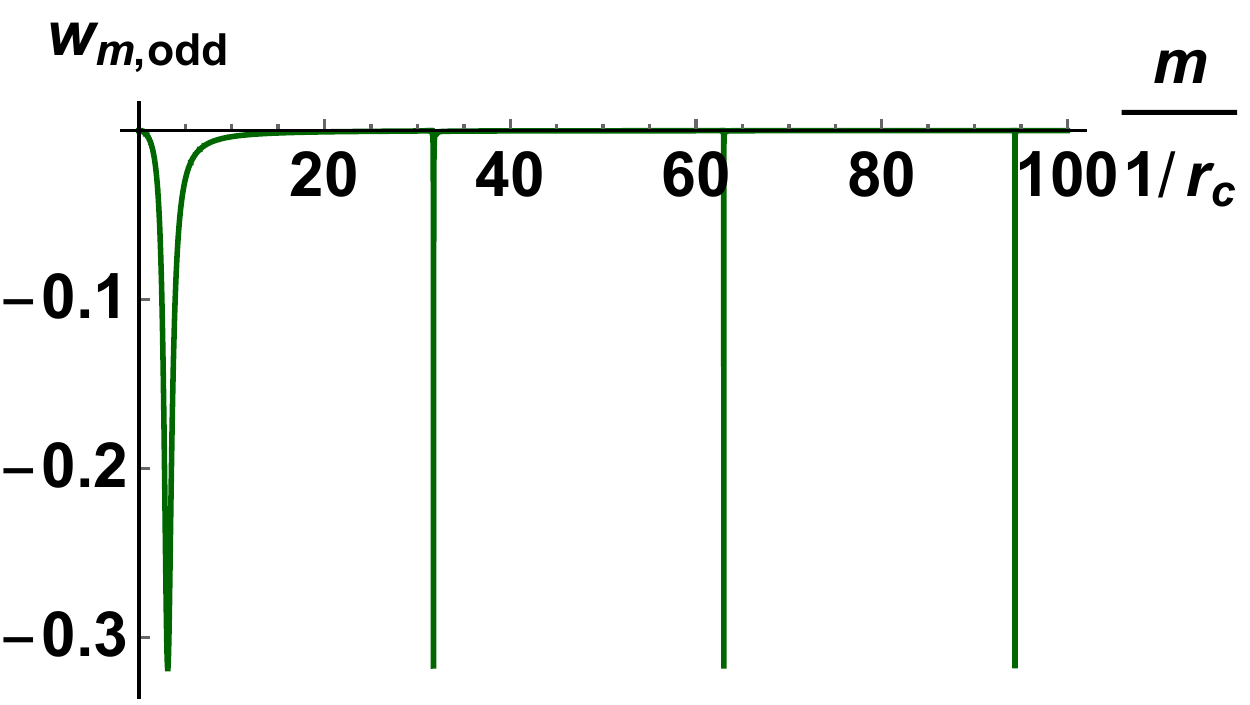}
        \caption{Odd}
    \end{subfigure}
    \caption{Wave profiles, with $m$ in units of $1/r_c$ and $R/r_c=0.2$}
    \label{fig:wave_profile}
\end{figure}

From the action (\ref{eq:KK_action}) we derive the equation of motion
\begin{equation}
\left( \Box + m^2 \right) \phi_{m,\alpha}(x^{\mu}) = J_{m,\alpha}(x^{\mu}) .
\end{equation}
Since we are interested in the static source
\begin{equation*}
J(x^{\mu},y)=g \left[ \delta^{(3)}(\vec{x}) \delta (y+R/2)+\delta^{(3)}(\vec{x}-\vec{r}) \delta (y-R/2) \right] ,
\end{equation*}
we have
\begin{equation}
J_{m,\alpha}(x^{\mu})=g \left[ \psi_{m,\alpha}(-R/2) \delta^{(3)}(\vec{x})+\psi_{m,\alpha}(R/2) \delta^{(3)}(\vec{x}-\vec{r}) \right] .
\end{equation}
Then the solution of
\begin{equation}
\left( - \Delta + m^2 \right) G_m(\vec{x}-\vec{x'}) =  \delta^{(3)}(\vec{x}-\vec{x'})
\end{equation}
leads to the Green's function
\begin{equation}
G_m(\vec{x}-\vec{x'}) = \int \frac{\dd^3 \vec{k}}{(2 \pi)^3}  \frac{\ee^{\iu \vec{k} (\vec{x}-\vec{x'})}}{|\vec{k}|^2+m^2}= \frac{1}{4 \pi r} \ee^{-mr} .
\end{equation}
Hence, the potential energy between the two sources is
\begin{equation}
V_{\text{KK}}(r,R) =- g^2 \sum_{\alpha} \int_0^{\infty} \dd m \, w_{m,\alpha} G_m(\vec{r}) , \label{eq:KK_potential}
\end{equation}
where we have defined the "wave profile" 
\begin{IEEEeqnarray}{rCl} 
w_{m,\alpha} &\equiv&  \psi_{m,\alpha}(-R/2) \psi_{m,\alpha}(R/2) \nonumber \\
&=& \left\{ \begin{aligned} 
&\frac{1}{\pi} \left[ 1+ \left(r_c m + \tan \left( \frac{m R}{2} \right) \right)^2 \right]^{-1},  & \al = \text{even}& ,  \\
& - \frac{1}{\pi} \left[ 1+ \left(r_c m - \cot \left( \frac{m R}{2} \right) \right)^2 \right]^{-1}, & \al = \text{odd}& , 
\end{aligned} \right.
\end{IEEEeqnarray}
which has been plotted in Figure \ref{fig:wave_profile}.

We observe that the wave profiles quickly tend to zero, except for distinct "peaks" (of height $1/\pi$) at certain values of $m$. The smaller $R/r_c$, the wider the peaks are apart. These distinct peaks come from the fact that the wave profiles have to satisfy certain matching conditions at the branes and it makes sense that for closer together branes there are less modes (or peaks) in a certain $m$-interval. So the modes, which contribute effectively to the potential, are discrete. For $m \gg \frac{1}{r_c}$ only those modes contribute, for which
\begin{IEEEeqnarray*}{rCl} 
\text{even:} \qquad \left| \tan{\frac{m R}{2}} \right| \gg 1 &\Rar& m \sim \frac{\pi}{R} (2 n -1) ,  \\
\text{odd:} \qquad \left| \cot{\frac{m R}{2}}  \right| \gg 1 &\Rar& m \sim \frac{2 \pi}{R} n ,
\end{IEEEeqnarray*}
with $n \in \N$.
Of course, the Yukawa-suppression factor determines which of those discrete modes eventually survive and contribute to the resulting potential. 

Another interesting observation is that the even wave profile always contributes positively (and hence to the attractive gravitational potential), while the odd wave profile always contributes negatively (and hence adds a repulsive part to the gravitational potential).

Now we want to see how these wave profiles conspire together, so that the potential turns out the same way as in the 5-d description. Unfortunately, in the KK-picture, it is very difficult to approximate the integral (\ref{eq:KK_potential}) in all, but the case (I).

We have to solve
\begin{equation}
V_{\text{KK}}(r,R) = - \frac{g^2}{4 \pi^2} \frac{1}{r r_c} \left( J_{\text{even}}-J_{\text{odd}} \right) ,
\end{equation} 
with
\begin{IEEEeqnarray}{rCl} 
\IEEEyesnumber \IEEEyessubnumber*
J_{\text{even}} &=&  \int_0^{\infty} \dd x \, \frac{\ee^{-\frac{r}{r_c} x}}{1+ \left[ x+\tan{\left( \hal \frac{R}{r_c} x \right)} \right]^2 } , \label{eq:J_even} \\
J_{\text{odd}} &=& \int_0^{\infty} \dd x \, \frac{\ee^{-\frac{r}{r_c} x}}{1+ \left[ x-\cot{\left( \hal \frac{R}{r_c} x \right)} \right]^2 } . \label{eq:J_odd}
\end{IEEEeqnarray}
Let us first consider case (I): Both integrals are cut off at $\frac{r_c}{r}$. The result of the integrals crucially depends on the interplay between the two terms in the square-bracket of the denominator. In $J_{\text{even}}$ the tangent is always sub-dominant, so
\begin{equation}
J_{\text{even}} \sim  \int_0^{\frac{r_c}{r}} \dd x \, \frac{1}{1+ x^2} = \arctan{\frac{r_c}{r}} \sim \frac{\pi}{2} - \frac{r}{r_c} + \mac{O} \left[ \left(\frac{r}{r_c} \right)^3 \right] .
\end{equation}
In  $J_{\text{odd}}$ the two terms in the square-bracket become of the same order for $x \sim x_* \equiv \sqrt{2} \frac{\rho}{R}$. But in the case (I), $x_* \gg \frac{r_c}{r}$, so the cotangent always dominates and we find
\begin{equation}
J_{\text{odd}} \sim  \frac{1}{4} \frac{R^2}{r_c^2} \int_0^{\frac{r_c}{r}} \dd x \, x^2 = \frac{1}{12} \frac{R}{r} \frac{\rho^2}{r^2} .
\end{equation}
Hence, the leading contribution to the potential comes from $J_{\text{even}}$ and we recover the same result (for the leading order) as in the 5-d description. 

Now, let us go to case (II). In $J_{\text{even}}$ the tangent is still sub-dominant. However, in $J_{\text{odd}}$ the story changes. Now $x_* \ll \frac{r_c}{r}$ and hence for $x \sim x_*$ the terms in the square-bracket cancel each other and the integrand has a finite contribution in that integration interval. We were not able to approximate the result analytically, but numerical calculations show that in the leading approximation (where we take $\ee^{-\frac{r}{r_c} x} \cong 1$) $J_{\text{odd}}$ goes to the same value as $J_{\text{even}}$. So, up to (at least) 2nd order the respective first peaks (see Figure \ref{fig:wave_profile}) cancel each other and the resulting potential energy vanishes. 

If we examine (\ref{eq:J_even}) and (\ref{eq:J_odd}) more closely (taking into account the Yukawa-suppression factor), we find that the first peak of $J_{\text{even}}$ and $J_{\text{odd}}$, respectively, together contribute the same leading factor $\frac{1}{\rho}$, which we found in the 5-d calculation. We can interpret this result as if those modes, which contribute to the repulsive potential (belonging to the first odd peak), counteract the modes, which contribute to the attractive potential (belonging to the first even peak), and thus weaken gravity. 

In case (III) the square-brackets in the denominators go to zero many times, so many of the peaks contribute to the result. A numerical analysis again shows that we recover the leading behavior of (\ref{eq:J_case_3}).

\section{Decoupling the second brane} \label{sec:decoupling}

We now want to consider two static point sources on our brane (but still in the presence of the parallel brane at $y=R$). The potential energy is
\begin{equation}
V_R (r) = - g^2 \int_0^{\infty} \frac{\dd^3 \vec{k}}{(2 \pi)^3} \, \ee^{\iu \vec{k} \vec{r}} \widetilde{G}(k,0) , \label{eq:potential_again}
\end{equation} 
where we can read off $\widetilde{G}(k,0)$ from (\ref{eq:propagator}). In the limit $R \to \infty$, the propagator becomes
\begin{equation}
\widetilde{G}(k,0) = \frac{1}{k} \frac{1}{2+r_c k} , \label{eq:approx_propagator}
\end{equation} 
which is the same as in the case with only one brane \cite{dvali20004d}. Hence,\footnote{If one doubts that (\ref{eq:approx_propagator}) is a good approximation, because values for $k \rar 0$ enter the integral in (\ref{eq:potential_again}), one can take the full propagator $\widetilde{G}(k,0)$ and approximate the resulting potential energy. We find that in the regime $r, r_c \ll R$ (for an arbitrary hierarchy between $r$ and $r_c$) the potential approaches indeed \eqref{eq:DGP_potential}. For the regime $r \ll R \ll r_c$ the Newtonian limit ($r \ll r_c$) of \eqref{eq:DGP_potential} is recovered.} for $R \to \infty$ we find the same result as in Ref. \cite{dvali20004d}:\footnote{The numerical prefactor can differ due to a different normalization (which can be absorbed in the coupling $g$).}
\begin{equation}
V(r) = - \frac{g^2}{2 \pi^2} \frac{1}{r r_c} \left\{ \sin{\left( \frac{2 r}{r_c} \right)} \mbox{Ci} \left( \frac{2 r}{r_c} \right) + \cos \left( \frac{2 r}{r_c} \right) \left[ \frac{\pi}{2} - \mbox{Si} \left(\frac{2 r}{r_c}\right)  \right] \right\} , \label{eq:DGP_potential}
\end{equation} 
where $\mbox{Si}(z)$ and $\mbox{Ci}(z)$ are the sine integral and the cosine integral, respectively. So we find that a 2nd brane together with its induced kinetic term has no influence on gravity in our world, if the brane is far away. 

In the previous discussion we used the same kinetic term on the two branes, by using the same length scale $r_c$ on them. However, we can reproduce the same conclusion, if we consider the theory
\begin{equation}
S = \int \mbox d^4 x \, \dd y \, \left\{ \hal \left( \del_A \phi \right)^2 + \left[ r_c  \delta \left(y \right) + \tilde{r}_c \delta \left(y-R \right) \right] \hal \left( \del_{\mu} \phi \right)^2 + J(x^{\mu},y) \phi \right\} ,
\end{equation}
where in general $ r_c \neq \tilde{r}_c$. We now find
\begin{equation}
\widetilde{G}(k,0) = \frac{1}{k} \frac{2+\tilde{r}_c k-\tilde{r}_c k \, \ee^{-2 k R}}{(2+\tilde{r}_c k) (2+r_c k) - r_c \tilde{r}_c k^2 \, \ee^{-2 k R}} ,
\end{equation} 
which again reduces to (\ref{eq:approx_propagator}) in the limit $R \rar \infty$.

\section{Consistency with black hole physics and interpretation in terms of species} \label{sec:BH_species}

In this last section we want to use our findings to discuss some implications for the present system, coming from the consistency with BH physics. Although the existence of BH's (and their exact form) in the DGP setup has not yet been conclusively established,\footnote{For some discussions see Ref. \cite{Gruzinov2001,Gabadadze2005}.} the following discussion will be based on the assumption that for distances $r \ll r_c$ there should exist BH's in the DGP model, which have the same properties as the usual 4-d BH's in GR.

\subsection{Classically static configuration}

In Ref. \cite{Dvali2008c} several scenarios (in the context of the ADD model) have been investigated, where a BH seemingly could not evaporate into the species localized on one or more branes, because its size was smaller than the distance to those branes. It has been found that all of those scenarios corresponded to time-dependent configurations. It was shown that a (classically) static configuration was only reached, after the BH has accreted all the branes (or had evaporated altogether) and henceforth could evaporate into all species. This was interpreted as a "democratic transition" from a "Non-Einsteinian" BH (meaning that it is not the usual time-independent, universally evaporating BH derived from GR) to a time-independent, semi-classical BH.

Now we want to apply this investigation to our setting. Let us consider a BH localized on "our" brane, in the presence of a parallel brane that has some species localized on it, which are not localized on our brane. In the situations of Ref. \cite{Dvali2008c}, the tension of the branes was the key property in deriving the gravitational interaction between the branes and the BH, and discovering the "accretion" mechanism. 

However, in our scenario the branes are tensionless, so one might naively suspect that they would not interact with a BH on our brane. But we found in Section \ref{sec:5_dim_potential} that the branes in our scenario, due to their DGP-term, would react to a present mass-source, such as a BH, by repelling it. So unless we treat the branes as fixed boundaries, our configuration would be again time-dependent. It goes beyond the scope of this paper to investigate, if a static configuration can be reached and to determine the time scale. However, the fact that the 2nd brane does not influence us, if it is very far away (see Section \ref{sec:decoupling}), suggests that there is at least the trivial static configuration, namely with the 2nd brane sent to infinity. In this case there could be a situation, where a BH on our brane cannot evaporate into the species on the parallel brane. 

This result differs from the result in the ADD case and it seemingly tells us that in the DGP scenario a BH that does not evaporate universally into all species does not have the tendency to "democratize" its evaporation. 

However, it is subtle to adopt such an interpretation. For the authors of Ref. \cite{Dvali2008c} to interpret the evolution towards time-independence (the accretion of the branes) as the evolution towards democratization, driven by the presence of the localized species on the distant branes, it was crucial to note the following relation: Since in a general coordinate invariant setup the branes have a dynamical origin (they are domain walls, instead of being merely boundary conditions), they must have a tension, which then will excite particles (even without additionally localized bulk fields). So an attraction between a BH and the branes is always accompanied by the presence of localized species on the latter. In our setup, however, the branes were originally introduced as boundary conditions and hence they can be tensionless. 

So our result either illustrates that in the DGP braneworld an interpretation in terms of species has its limits. Or the species picture tells us that our construction of two distant DGP branes is not consistent. For example, it might be the case that it is not possible to localize different species on the two branes in such a way.

\subsection{The dependence of the crossover scale on the number of species}

As explained in the introduction, in theories, which tend to GR in the infrared, consistency with BH physics requires a bound on the gravity cut-off. This bound depends on the number of species in the theory.\footnote{A similar relation as \eqref{eq:N_bound} can also be derived using the perturbative argument that the graviton propagator will get radiative corrections from the $N$ fields \cite{Dvali2001a,Veneziano2001}. However, we want to emphasize that the perturbative argument is in fact a naturalness argument, since the radiative corrections could cancel each other, whereas the non-perturbative (BH) argument leads to a consistency limit.} Now, in our scenario the situation is somewhat different. We consider a model, which goes to GR at short distances, but not at large distances. However, it is possible to derive a similar bound in our case, if one assumes the existence of the usual black holes. Furthermore, assuming the equality of the gravity cutoffs for a 5-d and a 4-d observer leads to an interesting relation between the crossover scale and the number of species.
In the following, we will reproduce the derivation, given in Ref. \cite{Dvali2009}.

If the existence of the usual 4-d BH with Schwarzschild radius $r_g=\frac{M}{M_*^3 r_c}=\frac{M}{M_{\te{P}}^2}$ for length scales $r < r_c$ is assumed, one can derive the same bound as in Eq. \eqref{eq:N_bound},
\begin{equation*}
\Lambda \lesssim  \frac{M_{\te{P}}}{\sqrt{N}} , 
\end{equation*} 
where $\Lambda$ is the cutoff of classical gravity and $N$ is the number of localized species on the brane. Let us, for definiteness, consider the case that we can have a semi-classical BH of a size all the way down to $\sqrt{N}/M_{\te{P}}$. Then, the highest possible cut-off will be
\begin{equation*}
\Lambda_{\text{max}}  = \frac{M_{\te{P}}}{\sqrt{N}} .
\end{equation*} 
Although in principle this gravity cutoff could be independent of the gravity cutoff for the 5-d bulk observer (who only sees one species), the authors in Ref. \cite{Dvali2009} explore the possibility that they are the same, so $\Lambda_{\text{max}} = M_*$. Since $r_c \equiv \frac{M_{\te{P}}^2}{M_*^3}$, this leads to
\begin{equation}
N = r_c M_* . \label{eq:N_DGP}
\end{equation} 

If we accept the relation (\ref{eq:N_DGP}), we see that, for fixed $M_*$ in the bulk, the number of localized particles is 
 \begin{equation}
N = \frac{M_{\te{P}}^2}{M_*^2} ,
\end{equation} 
and hence, from this point of view, is responsible for the strength of gravity on the brane. 

On the other hand, we saw in our discussion in the previous section that if the second brane is far enough away, then no matter how strong gravity is on that brane, it will not affect the gravitational laws in our world. So the above assumption of the equality of 5-d and 4-d gravity cutoffs leads to an interesting observation: even if there is a very high number of species in the theory (but localized on a different brane), it does not alter the gravity cutoff in our universe. This is again an implication, which differs from the situation in models, where there is a normalizable zero-mode of the graviton.

\section{Summary}

We have investigated the modification of the original DGP-setup \cite{dvali20004d} by the addition of a second 3-brane with a localized curvature term, parallel to the brane, where our SM fields are localized. We have shown that the potential energy between two static point masses shows similar properties as in the original DGP-setup, but it also acquires a qualitatively different behavior. By calculating the potential energy between two point masses on different branes, in the regime, where the cross-over scale $r_c$ is much bigger than the 4-dimensional distance $r$ and the brane separation $R$, we found that a new length scale emerges, namely $\rho = \sqrt{r_c R}$. For $\rho \ll r$, the (leading contribution of the) potential energy goes like $1/r$, while for $r \ll \rho$ it goes like $1/\sqrt{R}$. So we found that in the presence of the branes, 5-d gravity is screened. Although a screening was already found in Ref. \cite{Dvali2001b} for one brane, we established that the addition of the 2nd brane weakens gravity even further, so it behaves weaker (for $r \ll \rho$) than 4-dimensional.

Taking into account higher order contributions to the potential, we found a linear $r$-dependence in the regime $R \ll r \ll \rho$. Hence, if we consider the force along the 3-brane, there is a spatially constant attractive force between the point masses. This constant force, coming from matter sources on the parallel brane, can lead to rotation-velocities of objects, orbiting e.g. the center of our galaxy, which don't decrease, but rather slightly increase. This might be phenomenologically interesting for low surface brightness galaxies. Here, it might be worthwhile to consider extended sources on both branes and to investigate, in a future work, how the potential energy behaves and hence what rotation curves can be derived. 

We derived the potential energy in two different ways: we calculated it in 5 dimensions and derived an expression in terms of KK-modes after performing the dimensional reduction. In the KK-picture we showed that the even modes contribute only to the attractive potential, while the odd modes contribute only to the repulsive part, thus weakening gravity.

Finally, we have considered the limiting case, where the 2nd brane is sent far away ($R \gg r_c$). In this case, the laws of gravity on our brane behave like in the original DGP-scenario and are not influenced by the presence of another brane. If one assumes the existence of BH's at distances $r \ll r_c$ and adopts a species-viewpoint, this result is interesting for the following reasons. A BH on our brane will just evaporate into species localized on our brane. The strength of the gravitational potential and the gravity cut-off in our universe are not altered by the (possibly large) number of species on the distant brane. This suggests that either the usual bound on the number of species has limited applicability for a DGP-type model or that the species-viewpoint points to an inconsistency in our assumed construction of two DGP branes.

To gain more insight, one might investigate in a future work, how a classically static configuration (where the 2nd brane is at a finite distance) might look like in the presence of the branes and their repulsive nature. Would the BH also try to accrete the other brane?

\appendix

\section{Numerical demonstration of the validity of integral approximations} 

\subsection{Point sources on different branes} \label{sec:numerics_different}

\begin{figure}[t]
\centering
    \includegraphics[width=0.7\linewidth]{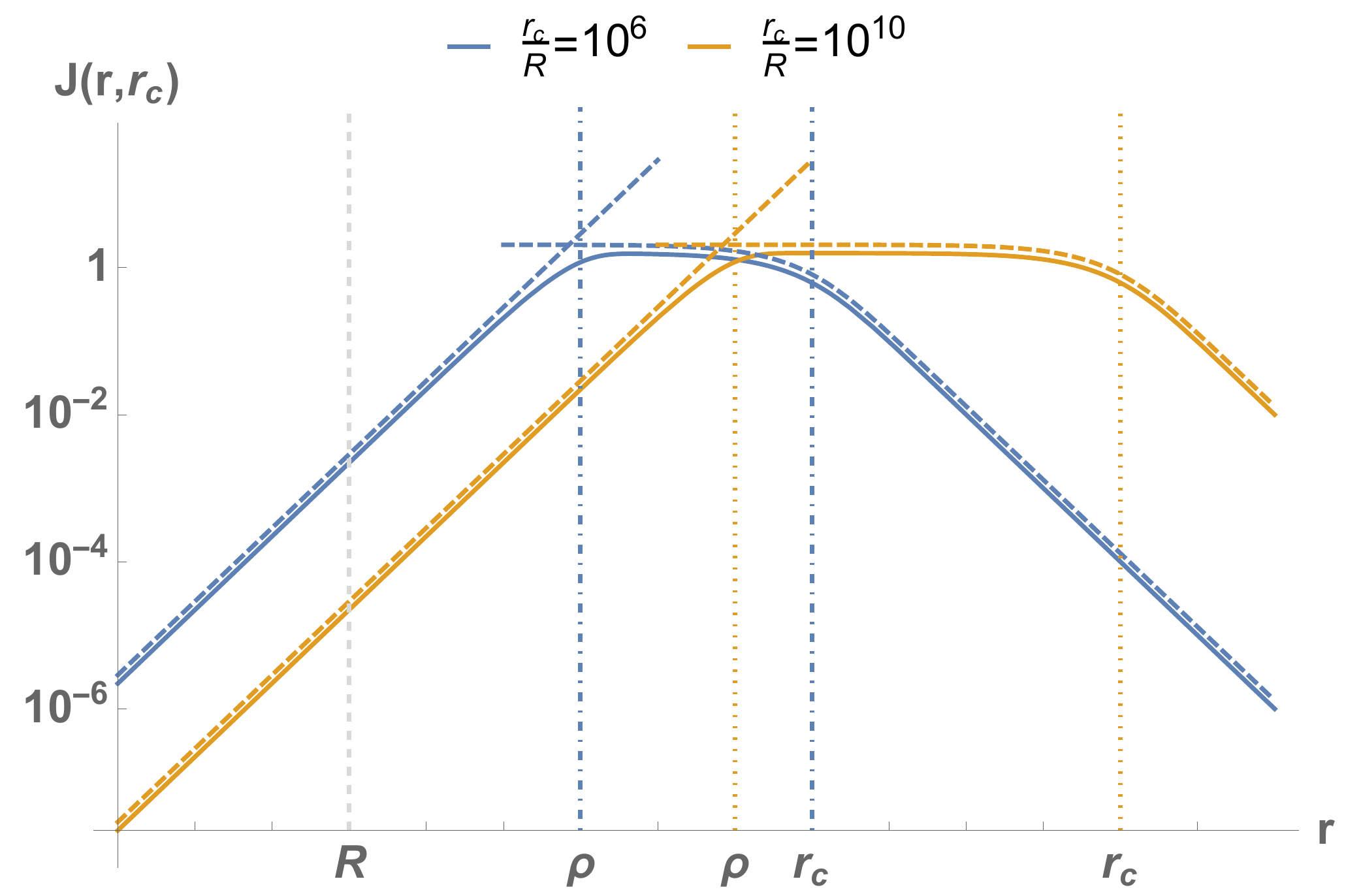}
    \caption[]{This log-log plot shows the numerical value of $J$, as a function of $r$ (in units of $R$), for two different values of $r_c/R$ as solid curves. The blue curve is plotted for $r_c/R=10^6$ and the orange one for $r_c/R=10^{10}$. The dashed curves are the functions $J_{\te{DGP}}$, for $r \gtrsim \rho$, and $\frac{\pi}{\sqrt{2}} \frac{r}{\rho}$, for $r \lesssim \rho$, in two different colors depending on the value of $r_c/R$. We see that $J$ is approximated very well by the two asymptotes in the two regimes, separated by $\rho$ (we have plotted the asymptotes with an offset, because otherwise the curves would lie on top of each other). The vertical lines show the scales, which separate the regimes, stated in Section \ref{sec:5_dim_potential}. We observe that $J$ does not change its leading order behavior by crossing $r \simeq R$. At $r \simeq \rho$, $J$ crosses over from a linear behavior to the "DGP behavior", where $J$ is approximately constant for $r \ll r_c$ and then crosses over to a $1/r$-behavior for $r \gg r_c$. The constant part is more pronounced for larger ratios $r_c/R$.}
    \label{fig:j_leading}
   \end{figure}

In order to show that the leading order approximations, which we found for the three specified regimes in Section \ref{sec:5_dim_potential}, are correct, we compare the numerical value of $J$ (obtained by numerical integration of \eqref{eq:J}), as a function of $r$, to the two asymptotes $J_{\te{DGP}}$ (from \eqref{eq:J_DGP}), valid for $r \gg \rho$, and $\frac{\pi}{\sqrt{2}} \frac{r}{\rho}$, valid for $r \ll \rho$. Since $J$ can be viewed as a function of the dimensionless variables $r/R$ and $r_c/R$, all functions, plotted in Figure~\ref{fig:j_leading}, depend just on $r$ and $r_c$, expressed in units of $R$.
We see that $J$ is approximated excellently by the given leading order approximations for the two regimes $r \ll \rho$ and $r \gg \rho$. Furthermore, if the ratio $\frac{r_c}{R}$ is large enough, $J$ can be approximated by a constant for $\rho \ll r \ll r_c$.

\begin{figure}[p]
\centering
    \includegraphics[width=0.7\linewidth]{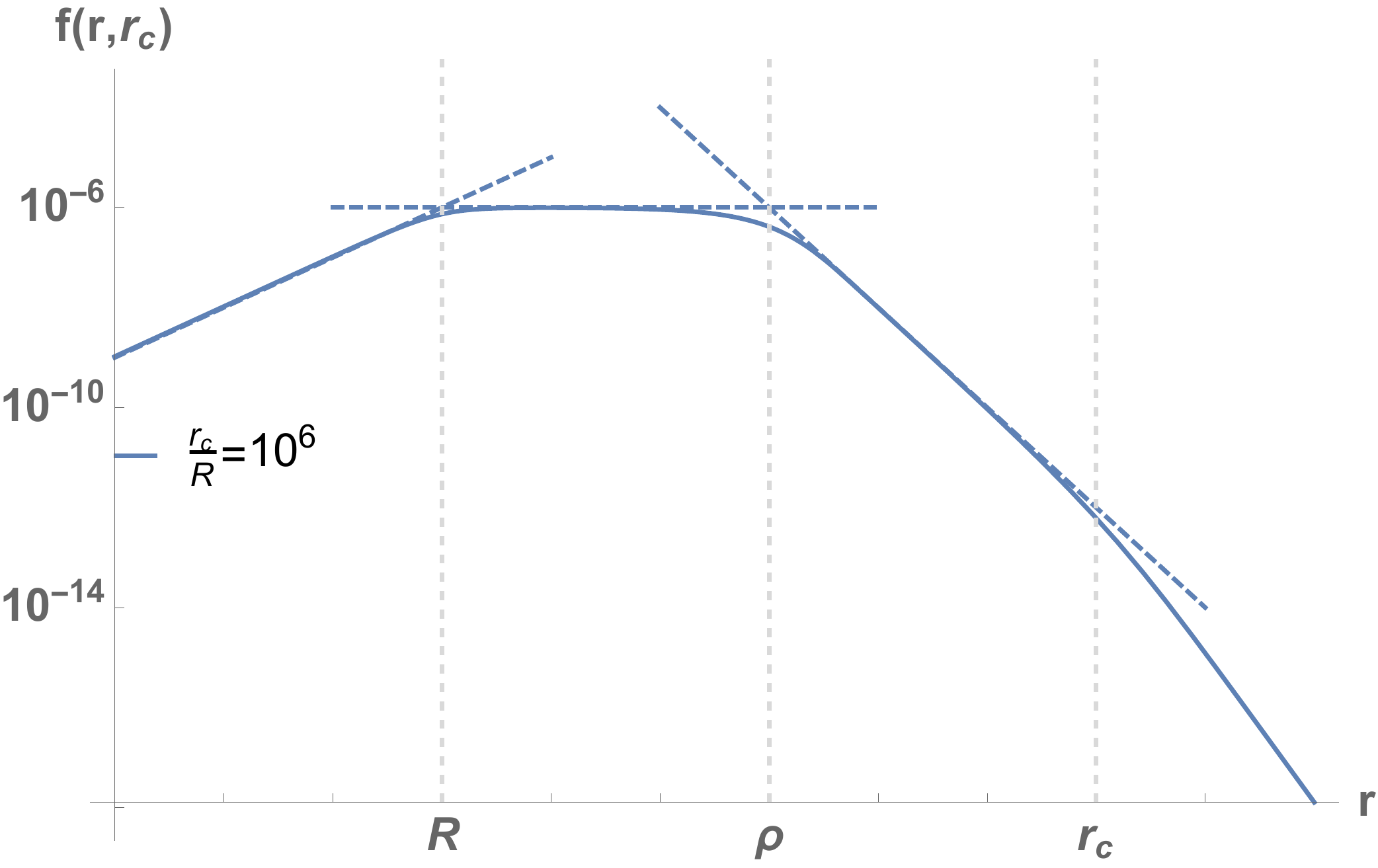}
    \caption{\label{fig:f_dark_leading} This log-log plot shows the numerical value of $f$, defined in \eqref{eq:f_numerical_int}, as a function of $r$ (in units of $R$) for the ratio $r_c/R=10^6$ as a solid curve. The dashed curves are the functions $\frac{r}{r_c}$ for $r \lesssim R$, $\frac{1}{r_c}$ for $R \lesssim r \lesssim \rho$ and $\frac{1}{r^2}$ for $r \gtrsim \rho$ (with again $r$ and $r_c$ in units of $R$). We see that $f$ is approximated very well by the three asymptotes in the three stated regimes. The dashed vertical lines show the scales, which separate the regimes, mentioned in Section \ref{sec:5_dim_potential}. We observe that already for $r_c/R=10^6$, the regime (II), where the force can be approximated as constant, extends over almost two orders of magnitude. Increasing $r_c/R$, extends this region. For $r \ll R$, the force vanishes linearly, while for $r \gg \rho$, it interpolates between a $1/r^2$- and a $1/r^3$-behavior.}

\vspace*{\floatsep}

  \includegraphics[width=0.7\linewidth]{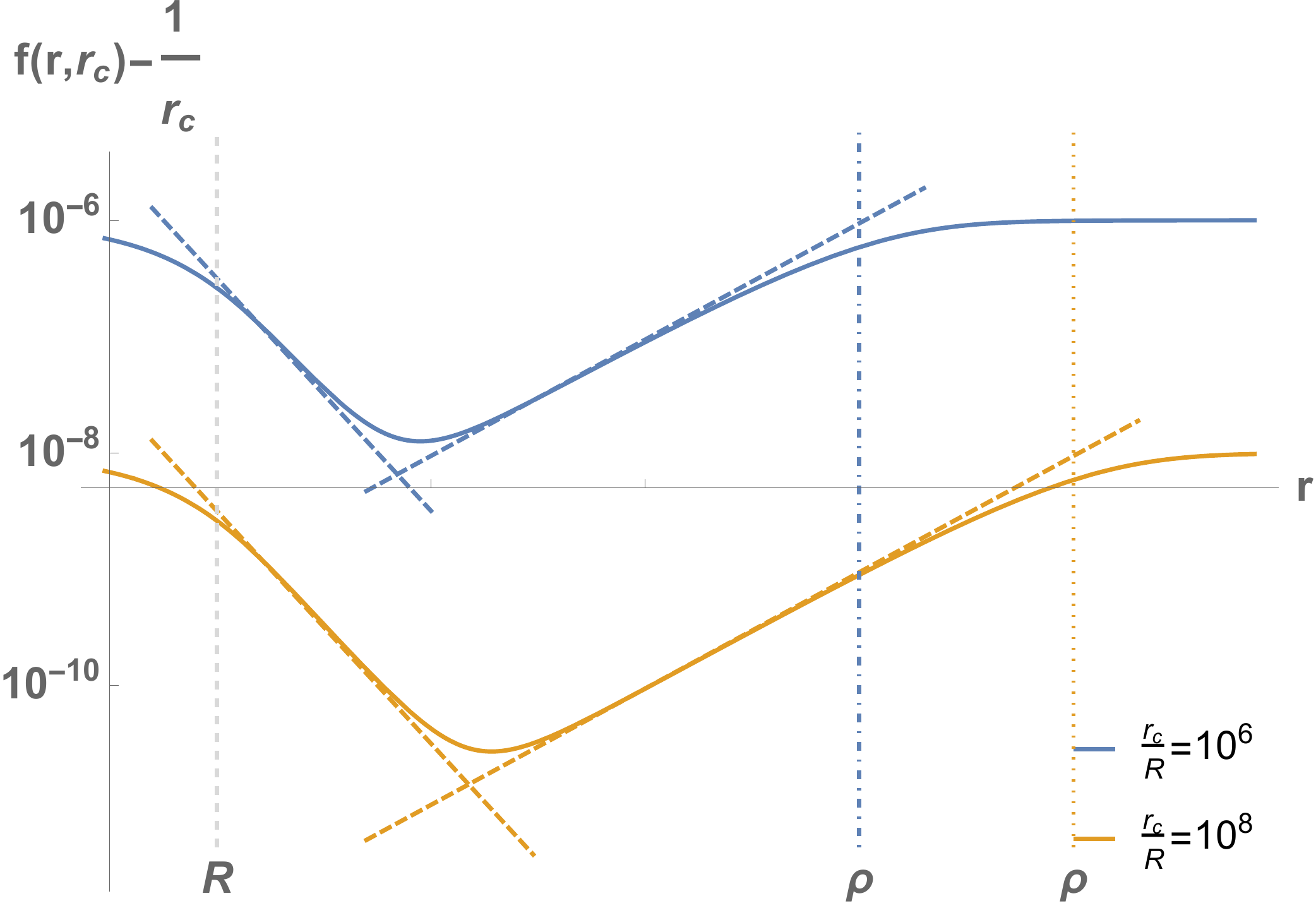}
    \caption[]{This log-log plot shows the numerical value of $f-\frac{1}{r_c}$ as a function of $r$ and $r_c$ (in units of $R$) for two different values of $r_c/R$ as solid curves. The blue curve is plotted for $r_c/R=10^6$ and the orange one for $r_c/R=10^8$. The dashed curves are the functions $\frac{1}{\pi r^2 r_c}$, for $r \rar R$, and $\frac{2 \sqrt{2}}{3} \frac{r}{r_c^{3/2}}$, for $r \rar \rho$, in two different colors depending on the value of $r_c/R$. We see that the corrections given in \eqref{eq:force} are good approximations for the numerical result, with $\frac{1}{\pi r^2} \frac{R}{r_c}$ dominating for smaller $r$ and $\frac{2 \sqrt{2}}{3} \frac{r}{\rho^3}$ dominating for larger $r$ in the regime (II). The vertical lines show the scales, which separate the regimes. We see that for larger $r_c/R$ the approximation gets better.}
    \label{fig:f_dark_correction_2}
   \end{figure}

Next, let us verify that also the corrections to the leading order terms of $J$, given in Section \ref{sec:5_dim_potential}, are accurate. Since in Section \ref{sec:5_dim_potential} we are interested mostly in the force along the brane, where the terms linear in $r$ will drop out, in the remainder of this subsection, we will perform our numerical analysis exclusively for the force \eqref{eq:force}. In order to plot a dimensionless quantity $f$, we define the physical force as 
\begin{equation}
F_r \equiv - \frac{1}{16 \pi M_{\te{P}}^2} \frac{1}{R^2} f(r,r_c) \equiv - \frac{G}{2} \frac{1}{R^2} f(r,r_c) , \label{eq:f_numerical}
\end{equation}
so that\footnote{\label{fn:numerical_force} In order to extract a numerical result, we pull the derivative inside the integral of \eqref{eq:J} and integrate numerically.}
\begin{equation}
f \equiv - \frac{2}{\pi} R^2 \frac{\partial}{\partial r} \left( \frac{J}{r} \right) . \label{eq:f_numerical_int}
\end{equation}
We again consider $f$ as a function of $r$ and $r_c$ (in units of $R$). Figure~\ref{fig:f_dark_leading}
shows the force along the brane between two point sources on different branes. We see that the leading order terms of \eqref{eq:force} approximate the actual behavior of the force very accurately in the regimes (I)-(III), given in Section \ref{sec:5_dim_potential}.

To justify the corrections to the leading order terms in the expression \eqref{eq:force}, let us consider regimes (I) and (II) separately. For $R \ll r \ll \rho$, we compare the numerical value of $f$, where we subtracted the leading order term $\frac{R}{r_c}$, to the correction terms given in \eqref{eq:force} (the numerical prefactors are specified in the caption of Figure~\ref{fig:f_dark_correction_2}), in Figure~\ref{fig:f_dark_correction_2}.
    
We see that the two given corrections in \eqref{eq:force} are in good agreement with the numerical result, where $\frac{1}{\pi r^2} \frac{R}{r_c}$ dominates for smaller $r$, while $\frac{2 \sqrt{2}}{3} \frac{r}{\rho^3}$ dominates for larger $r$, in the regime (II). 

\begin{figure}[t]
\centering
    \includegraphics[width=0.7\linewidth]{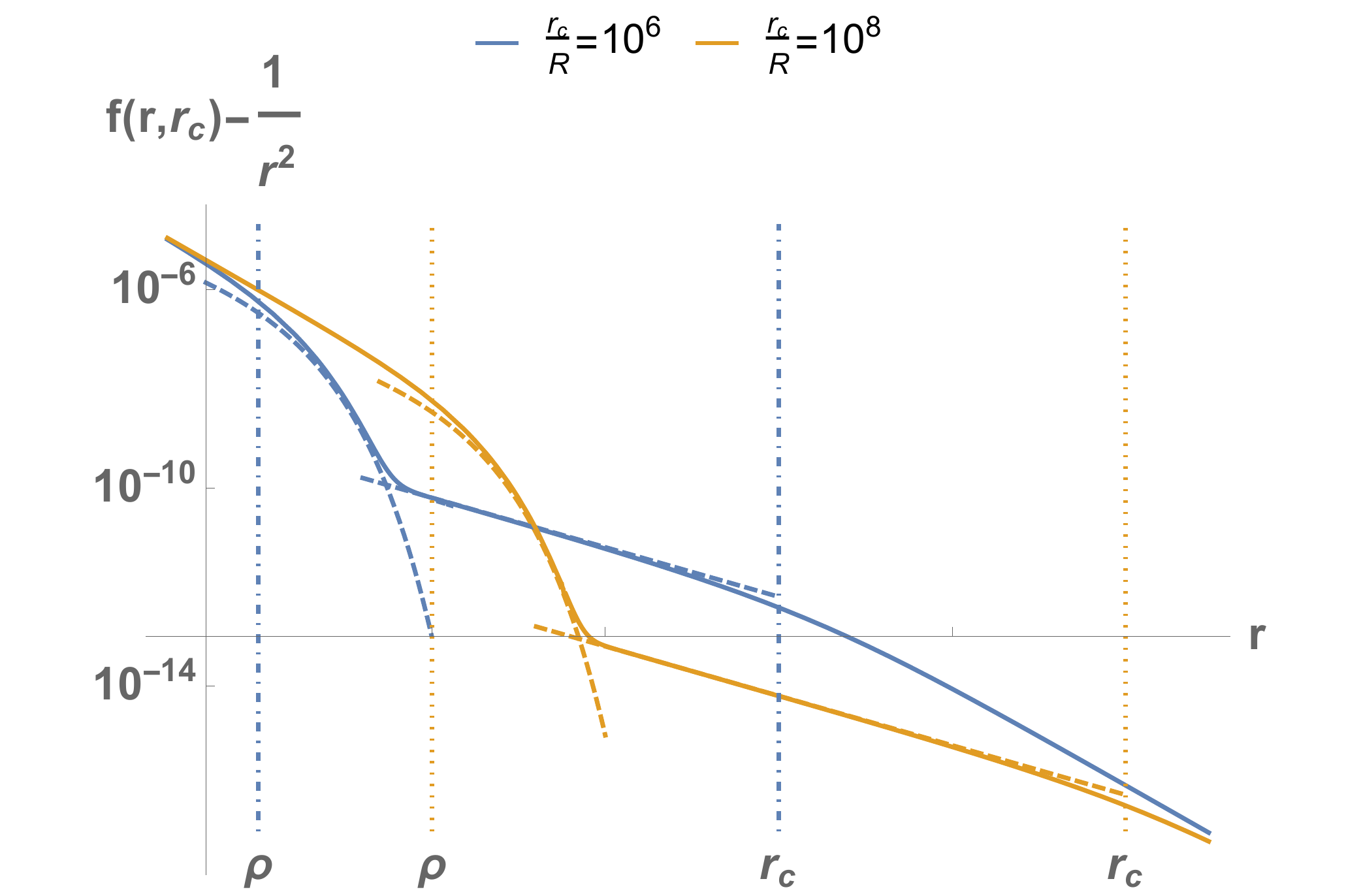}
    \caption[]{This log-log plot shows the numerical value of $f-\frac{1}{r^2}$ as a function of $r$ (in units of $R$) for two different values of $r_c/R$ as solid curves. The blue curve is plotted for $r_c/R=10^6$ and the orange one for $r_c/R=10^8$. The dashed curves are the functions $\frac{\sqrt{2}}{r \sqrt{r_c}} \ee^{-\sqrt{2} \frac{r}{\sqrt{r_c}}}$, for $r \rar \rho$, and $\frac{2}{\pi} \frac{1}{r r_c}$ for $r \rar r_c$ (with $r$ and $r_c$ in units of $R$), again for two different values of $r_c/R$. We see that the corrections given in \eqref{eq:force} are good approximations for the numerical result, with $\frac{\sqrt{2}}{r \rho} \ee^{-\sqrt{2} \frac{r}{\rho}}$ dominating for smaller $r$ and $\frac{2}{\pi} \frac{1}{r r_c}$ dominating for larger $r$, in the regime (I). The vertical lines show the scales, which separate the regimes. We see that for larger $r_c/R$ the approximation gets better.}
    \label{fig:f_dark_correction_1}
   \end{figure}

Next, we compare the corrections for $\rho \ll r$, given in \eqref{eq:force}, to the numerical value of $f$, where we subtracted the leading order term $\frac{R^2}{r^2}$. We show in Figure~\ref{fig:f_dark_correction_1}
that the two corrections, given in \eqref{eq:force}, are again quite good approximations for the numerical result, where the correction $\frac{\sqrt{2}}{r \rho} \ee^{-\sqrt{2} \frac{r}{\rho}}$ dominates for $r \rar \rho$, while the correction $\frac{2}{\pi} \frac{1}{r r_c}$ dominates for $r \rar r_c$.

\subsection{Point sources on the same brane} \label{sec:numerics_same}

\begin{figure}[t]
\centering
    \includegraphics[width=0.7\linewidth]{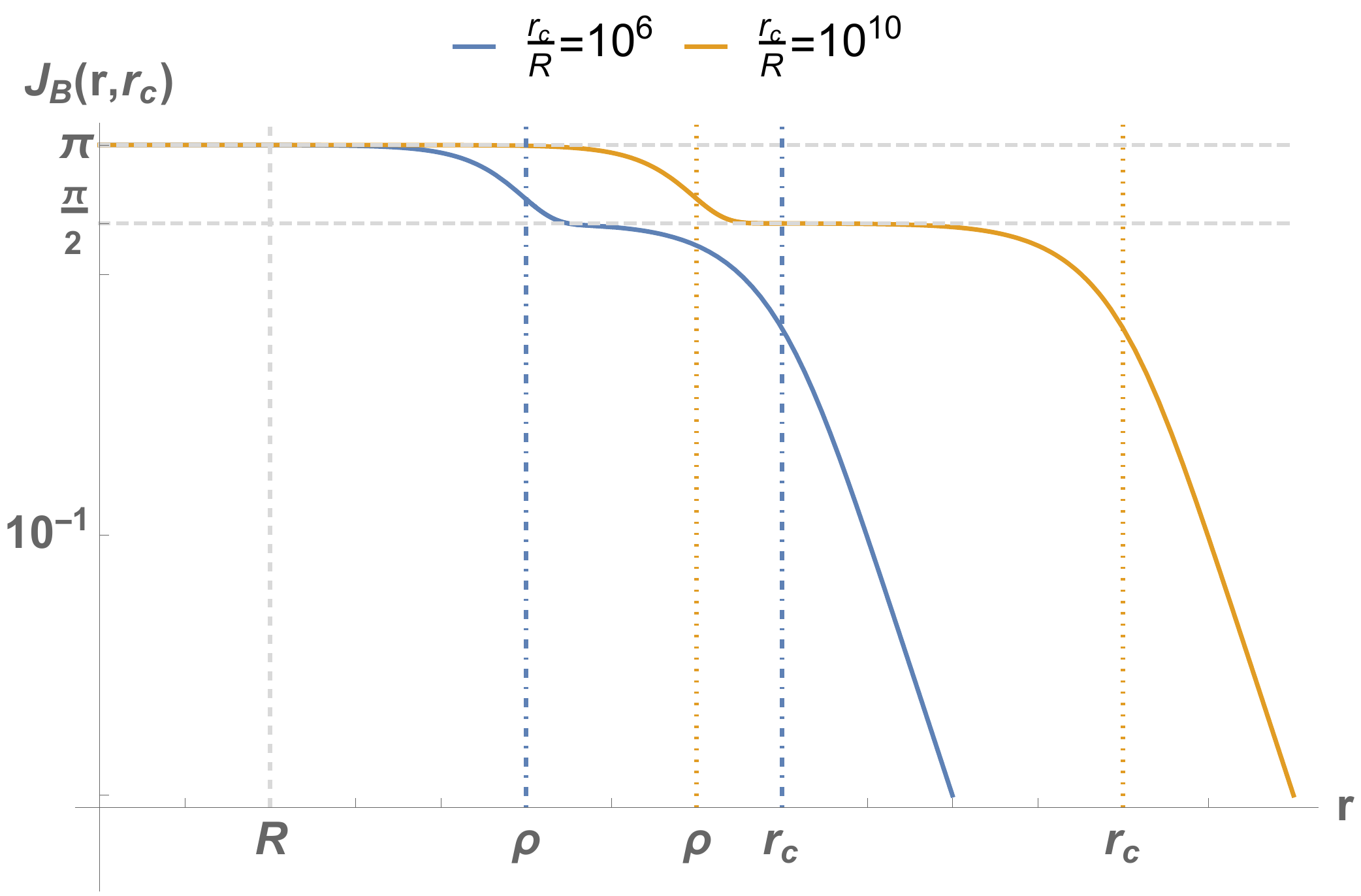}
    \caption[]{This log-log plot shows the numerical value of $J_{\te{B}}$ as a function of $r$ (in units of $R$) for two different values of $r_c/R$ as solid curves. The blue curve is plotted for $r_c/R=10^6$ and the orange one for $r_c/R=10^{10}$. The dashed horizontal lines go through the points $\pi$ and $\pi/2$. We see that $J_{\te{B}}$ starts with a constant value of $\pi$ for $r \ll \rho$, does not change across $R$ and then crosses over to a DGP-behavior at $r \sim \rho$, where it goes from a constant value of $\pi/2$ to a $\frac{1}{r}$-behavior. Again, the constant region ($\pi/2$) is more pronounced for larger values of $r_c/R$. The vertical lines show the scales, which separate the regimes, stated in Section~\ref{sec:5_dim_potential}.}
    \label{fig:f_baryonic}
   \end{figure}

The potential energy due to two point sources on "our" brane in the presence of an empty second DGP brane, is given by \eqref{eq:potential_again} with \eqref{eq:propagator}. This leads to the "baryonic" potential energy
\begin{equation}
V_{\te{B}}(r) = - \frac{g^2}{4 \pi^2} \frac{1}{r r_c} J_{\te{B}} , \label{eq:potential_baryonic}
\end{equation} 
with
\begin{equation}
J_{\te{B}} = 2 \int_0^{\infty} \dd x \, \sin{x} \frac{2 \frac{r}{r_c}+x-x \, \ee^{-2\frac{R}{r} x}}{\left(2 \frac{r}{r_c}+x \right)^2 - x^2 \, \ee^{-2\frac{R}{r} x}} . \label{eq:J_B}
\end{equation} 
The plot in Figure~\ref{fig:f_baryonic}
shows the numerical result for \eqref{eq:J_B} for two different values of $\frac{R}{r_c}$. We see that for $r \ll \rho$, $J_{\te{B}}=\pi$, so we recover the usual Newton's potential. In particular, there is no change of behavior in going from $r \lesssim R$ to $r \gtrsim R$. Only when $r$ approaches $\rho$ does the potential energy crosses over to the DGP-behavior, found in Ref. \cite{dvali20004d}. So if $r_c/R$ is large enough, there will still be a region, where the potential is approximately Newtonian, but with half its value (see footnote~\ref{fn:half_potential_energy}). One of the results of the present work is that we found the length scale, which differentiates between these two behaviors, to be $\rho$.

\section{Determining the mode functions} \label{sec:mode_functions}

For every $m$ there are two linearly independent solutions, which take the form
\begin{equation}
\psi_m(y) = \left\{ 
\begin{aligned} 
&A_m \cos{m y} + B_m \sin{m y} , & y<-\frac{R}{2}& , \\ 
&C_m \cos{m y} + D_m \sin{m y} , \quad & -\frac{R}{2} \le y \le \frac{R}{2}& , \\
&E_m \cos{m y} + F_m \sin{m y} ,  & y>\frac{R}{2}&  ,
\end{aligned} \right. 
\end{equation}
subject to the four boundary conditions: continuity of $\psi_m(y)$ at $y=-\frac{R}{2}$ and $y=\frac{R}{2}$, discontinuity of $\frac{\dd \psi_m(y)}{\dd y}$ at $y=-\frac{R}{2}$ and $y=\frac{R}{2}$.\footnote{$\lim \limits_{\epsilon \to 0} \frac{\dd \psi_m(y)}{\dd y}\Bigr|^{\pm \frac{R}{2} +\epsilon}_{\pm \frac{R}{2}-\epsilon}+m^2 r_c \psi_m(\pm \frac{R}{2})=0$.} This enables us to eliminate four of the coefficients and find
\begin{align*}
A_m &= C_m \left(1+r_c m \sin{\left(\frac{m R}{2}\right)} \cos{\left(\frac{m R}{2}\right)} \right) -r_c m D_m \sin^2{\left(\frac{m R}{2}\right)}, \\ 
B_m &= D_m \left(1-r_c m \sin{\left(\frac{m R}{2}\right)} \cos{\left(\frac{m R}{2}\right)} \right) +r_c m C_m \cos^2{\left(\frac{m R}{2}\right)}, \\ 
E_m &= C_m \left(1+r_c m \sin{\left(\frac{m R}{2}\right)} \cos{\left(\frac{m R}{2}\right)} \right) +r_c m D_m \sin^2{\left(\frac{m R}{2}\right)}, \\
F_m &= D_m \left(1-r_c m \sin{\left(\frac{m R}{2}\right)} \cos{\left(\frac{m R}{2}\right)} \right) -r_c m C_m \cos^2{\left(\frac{m R}{2}\right)} .
\end{align*}
In principle, every choice of $C_m$ and $D_m$ parametrizes a different solution in the 2-dimensional solution-space. However, since the operator acting on $\psi_{m,\alpha}(y)$ in eq. (\ref{eq:schroedinger}) is symmetric around $y=0$, we can find solutions that are either even ($\alpha=1$) or odd ($\alpha=2$). Hence, our mode functions are
\begin{equation}
\psi_{m,1}(y) \equiv \psi_{m,\text{even}}(y)  = \left\{ \begin{aligned} 
&A_m \cos{m y} + B_m \sin{m y} , \quad & y<-\frac{R}{2}& , \\ 
&C_m \cos{m y} , \quad & -\frac{R}{2} \le y \le \frac{R}{2}& , \\
&A_m \cos{m y} -B_m \sin{m y} , \quad & y>\frac{R}{2}& ,
\end{aligned} \right. 
\end{equation}
with $A_m = C_m \left(1+r_c m \sin{\left(\frac{m R}{2}\right)} \cos{\left(\frac{m R}{2}\right)} \right)$, $B_m = r_c m C_m \cos^2{\left(\frac{m R}{2}\right)}$ and
\begin{equation}
\psi_{m,2}(y) \equiv \psi_{m,\text{odd}}(y)  = \left\{ \begin{aligned} 
&A_m \cos{m y} + B_m \sin{m y} , \quad & y<-\frac{R}{2}& , \\ 
&D_m \sin{m y} , \quad & -\frac{R}{2} \le y \le \frac{R}{2}& , \\
&-A_m \cos{m y} + B_m \sin{m y} , \quad & y>\frac{R}{2}& ,
\end{aligned} \right.
\end{equation}
with $A_m = -r_c m D_m \sin^2{\left(\frac{m R}{2}\right)}$, $B_m = D_m \left(1-r_c m \sin{\left(\frac{m R}{2}\right)} \cos{\left(\frac{m R}{2}\right)} \right)$. It is straightforward to establish that these mode functions satisfy condition (\ref{eq:completeness}).

Performing that lengthy calculation, fixes the coefficients
\begin{align}
C_m^2 &= \frac{1}{\pi} \frac{1}{\cos^2{\left( \frac{m R}{2}\right)}+{\left( \sin{\left( \frac{m R}{2}\right)}+r_c m \cos{\left( \frac{m R}{2}\right)}\right)}^2} , \\
D_m^2 &= \frac{1}{\pi} \frac{1}{\sin^2{\left( \frac{m R}{2}\right)}+{\left( \cos{\left( \frac{m R}{2}\right)}-r_c m \sin{\left( \frac{m R}{2}\right)}\right)}^2} .
\end{align}

\acknowledgments

We are very grateful to Gia Dvali for suggesting this project to us and thank him for valuable discussions and comments. Further, we thank Lasha Berezhiani for discussions and useful comments, in particular on the part about dark matter. We also want to thank the anonymous referee, who, upon submission of the manuscript to the \emph{Journal of High Energy Physics}, pointed out some weak points in our presentation of the approximation of the potential energy. This led to a much more extended demonstration of the validity of those approximations. A special thanks is due to Lukas Eisemann, who, restlessly, supported us in the search of finding better approximations and created most of the numerical plots, shown in this work. This work was supported by the ERC Advanced Grant 339169 "Selfcompletion".

\bibliographystyle{JHEP}
\bibliography{library_phd_thesis}

\end{document}